\documentclass[]{aastex63}             %
\usepackage{amsmath}

\newcommand\Hzs{Hz~s$^{-1}$ }
\newcommand\Hzsns{Hz~s$^{-1}$}

\received{July 28, 2021}
\revised{October 25, 2021}
\accepted{December 7, 2021}

\submitjournal{Astronomical Journal}

\shorttitle{An ML-Based Filter for RFI Identification in the Search for Technosignatures}
\shortauthors{Pinchuk \& Margot}

\usepackage{hyperref}
\hypersetup{colorlinks=true,citecolor=blue,linkcolor=blue,urlcolor=blue,breaklinks=true}

\begin{document}

\title{A Machine Learning-based Direction-of-origin Filter for the Identification of Radio Frequency Interference in the Search for Technosignatures}

\correspondingauthor{Jean-Luc Margot}
\email{jlm@epss.ucla.edu}

\author[0000-0003-4736-4728]{Pavlo Pinchuk}
\affiliation{Department of Physics and Astronomy, University of California, Los Angeles, CA 90095, USA}

\author[0000-0001-9798-1797]{Jean-Luc Margot}
\affiliation{Department of Earth, Planetary, and Space Sciences, University of California, Los Angeles, CA 90095, USA}
\affiliation{Department of Physics and Astronomy, University of California, Los Angeles, CA 90095, USA}

\begin{abstract}
Radio frequency interference (RFI) mitigation remains a major
challenge in the search for radio technosignatures.  Typical
mitigation strategies include a direction-of-origin (DoO) filter,
where a signal is classified as RFI if it is detected in multiple
directions on the sky.  These classifications generally rely on
estimates of signal properties, such as frequency and frequency drift
rate.  Convolutional neural networks (CNNs) offer a promising
complement to existing filters because they can be trained to analyze
dynamic spectra directly, instead of relying on inferred signal
properties. In this work, we compiled several data sets consisting of
labeled pairs of images of dynamic spectra, and we designed and
trained a CNN that can determine whether or not a signal detected in
one scan is also present in another scan.  This CNN-based DoO filter
outperforms both a baseline 2D correlation model as well as existing
DoO filters over a range of metrics,
with precision and recall values of 99.15\% and 97.81\%, respectively.
We found that the CNN reduces the number of signals requiring visual
inspection after the application of traditional DoO filters by a
factor of
6--16 in nominal situations.
\end{abstract}

\section{Introduction} \label{sec:intro}

Radio technosignature searches have increased dramatically both in scope and
complexity since the early days of the search for extraterrestrial
intelligence \citep[][and references therein]{drak65, Tarter2001, Tarter2010, drak11}. 
In the past three years alone, the UCLA SETI Group's radio 
technosignature detection algorithms have undergone multiple levels 
of improvements, increasing the total number of detections in a typical
2-hour observing window by more than an order of magnitude 
\citep{Pinchuk2019, Margot2021}. 
Our current pipeline detects $\sim$200 times more 
signals per unit bandwidth per unit integration time than recent Breakthrough 
Listen (BL) searches \citep{Enriquez2017, Price2020, gajj21}
\added{with the same telescope, receiver, and detection threshold}.
We have also improved our
radio frequency interference (RFI) excision 
algorithms, yielding
RFI classification accuracies of $>99\%$ on data sets with 
millions of candidate signals.
Other groups are making progress along these lines as well.
For instance, the custom hardware system that enabled the early
1990s NASA High Resolution Microwave Survey was migrated to a software
platform whose RFI excision capabilities have continued to evolve~\citep{Harp2016}.
\citet{traas21} reported the results from a
search of 28 targets selected from the \textit{TESS} Input Catalog 
and also described an improvement to the BL RFI excision technique.

Despite these advancements, RFI remains the biggest challenge to the 
search for technosignatures.
\citet{Pinchuk2019} and \citet{Margot2021}
described several pitfalls of
current RFI identification algorithms that rely on
inferred signal properties,
such as estimates of frequency and frequency drift rate. 
They 
suggested that
these hurdles
might be overcome by an algorithm that instead examines the structure
of candidate signals in time-frequency space.  Because the
time-frequency structure of a signal resembles an image, we can
readily apply modern computer vision techniques to this problem, as
also suggested by \citet{Cox2018}, \citet{Zhang2018}, \citet{Harp2019},
and \citet{BLML2020}.

The last decade (2010-2020) has seen considerable advances in the field of
Convolutional Neural Networks (CNNs). In 2012, \citet{alexnet} 
introduced the AlexNet architecture, which won the ImageNet ILSVRC challenge 
\citep{ILSVRC15} the same year.
This architecture achieved a top-five error rate of $17\%$, which represents 
the percentage of test images (256$\times$256 pixels) for which the network's top 
five predictions, chosen from a total of 1000 classes, did not include the 
correct answer. This was an unprecedented accomplishment at the time.
An explosion of CNN architectures followed in subsequent years, each 
performing better than the last \citep{vgg, googlenet, resnet, xception}. 
Modern CNN architectures have achieved top-five error rates of $3\%$ or less 
\citep[e.g.,][]{tan2019efficientnet}. 

Machine learning (ML) has permeated both the workforce and the research industry, 
often leading to large improvements in challenging classification problems.
In particular, astronomers have already applied CNNs
to push the boundaries of astronomical data analysis. For example, 
\citet{schawinski2017} trained a Generative Adversarial Network composed of 
two CNNs (one to classify samples and another to generate them during 
training) to recover features such as galaxy morphology from 
low signal-to-noise ratio (S/N) and low angular resolution images.
\citet{shallue2018} trained a deep CNN to predict 
whether a signal found in \textit{Kepler} data was a transiting exoplanet or 
a false positive, allowing them to detect
and validate a five--planet resonant chain around Kepler--80 and a new, 
eighth planet around Kepler--90. \citet{zhang2018FRB} detected 72 new pulses 
from the repeating fast radio burst FRB 121102 using a
CNN trained on radio astronomy data obtained with the Green Bank 
Telescope. For a more detailed overview of ML and CNNs 
applied to astronomy, see \citet{baron2019} and references therein.

CNN applications
have been explored in the context of radio 
technosignature searches. \citet{Cox2018} and \citet{Harp2019} both 
generated a labeled set of synthetic candidate signals from a small 
($<10$) number of RFI classes to train a CNN for RFI classification. Although
this approach provides a relatively simple way to obtain a labeled training set, 
the synthetic signals may not be representative enough of
actual signals and may therefore introduce
a
bias during model
training. \citet{Zhang2018} avoided this problem by using
self-supervised learning to train their network. Specifically, their
CNN was trained to predict the future time-frequency structure of a
signal given the time-frequency structure from a past subset of the
total observation. This method allowed the observations to act as both the
training set and the training labels for the model. However, in order
to apply the network for RFI excision, the similarity of 
the predicted signals and the observed signals must be evaluated.
Because this task is not trivial, the RFI classification performance of the network may suffer.
\citet{BLML2020} explored the application of CNNs
to technosignature candidate signal detection. Specifically, the
authors trained a CNN to detect up to two signals in
a frequency span of
$\sim$1400 Hz.  Although a potential improvement over the detection
algorithms of \citet{Enriquez2017} and \citet{Price2020}, this CNN
cannot yet compete with the detection algorithms described by
\citet{Margot2021}, which can detect hundreds of signals within the
same frequency range.

In this work, we describe an %
application of CNNs to the 
excision of RFI in technosignature data. The article is organized as follows.  
Our motivation and approach to this problem are presented in 
Section~\ref{sec:appr}. Our data compilation procedure is detailed in 
Section~\ref{sec:data_prep}.
In Section~\ref{sec:models}, we describe 
our approach to 
CNN model selection and hyperparameter tuning.
We also describe a non-ML baseline model that we use as a point of
comparison to the trained CNN.  Section~\ref{sec:results} summarizes
our results, including the final model performance on the test set as
well as on archival data.  In Section~\ref{sec:discussion}, we
describe several failure modes of our trained network and offer
avenues for future improvements.  We
present conclusions in
Section~\ref{sec:conclusions}.

\section{Motivation and Approach} \label{sec:appr}
Modern radio technosignature programs detect millions of signals
per survey \citep[e.g.,][]{Siemion2013, Harp2016, Enriquez2017, Margot2018, Pinchuk2019, 
  Price2020, Margot2021, gajj21}.
These signals must be carefully analyzed to 
determine whether or not they are of anthropogenic nature. The standard approach to perform this analysis is the direction-of-origin (DoO) filter. 
This filter labels a signal as RFI if it is not 
persistent in one direction
on the sky or if it is detected in multiple 
directions
on the sky.
Theoretically, this filter is powerful enough to 
remove all RFI signals that are detected in multiple scans.
In practice, 
this filter often fails on a
small 
subset of signals,
but even failure rates as low as $1\%$ can be costly because visual inspection
of the remaining signals may be necessary.  For instance, the filter
failure rates in the searches of \citet{Pinchuk2019} and
\citet{Margot2021} were $1.66\%$ and $0.162\%$, respectively,
requiring further examination of 96,940 and 43,020 signals,
respectively.

The main pitfall of the DoO filter is the accuracy
with which a 
unique signal can be linked across multiple scans.
This ``signal pairing'' is required for both the
persistence-test
(present in all scans of the source) and the uniqueness-test (absent in scans of other 
sources) portions of the filter. Different surveys implement this pairing
functionality in various ways. For example,
\citet{Enriquez2017} and \citet{Price2020} considered two signals to be from a 
common origin if the frequency at which the latter signal is detected is 
within a generous tolerance of $\pm 600$Hz of the detection frequency of the 
first signal,
even if the corresponding frequency drift rates are unrelated.
Although this
approach speeds up the analysis by discarding 
a large portion of the candidate signals,
it is problematic because
it may eliminate valid technosignatures.
A more
rigorous
approach was adopted by
\citet{Pinchuk2019} and \citet{Margot2021}.
In both of these searches,  two
signals were paired only if their frequency drift rates and
frequencies extrapolated to a common epoch are within a small
tolerance.
More robust versions of this filter could include
tests of other
signal properties, such as signal bandwidth or off-axis gain ratio.

In all four searches described in the preceding paragraph, the filter was applied to
estimates of signal
properties
produced by a computer program on the basis of the 
time-frequency structure of each signal. Therefore, the efficiency of
the filter relies heavily on the accuracy of the derived signal
properties.
When the estimates of these signal properties are imprecise or
incorrect, or when the underlying assumption of a linear drift rate is
violated, the filter classification fails. \citet{Pinchuk2019}
detailed five different signal types for which their
DoO filter exhibited a degraded performance.
Importantly, this limitation can
likely
be
overcome by
an algorithm that examines the time-frequency
structure of each signal directly.  %

In this work, our approach is to train a CNN to pair signals by
directly examining the corresponding dynamic spectra.
The trained network is then used to examine the data as follows.
For each signal detected in the survey, we extract
a portion of the
dynamic spectrum centered on the time-frequency location of the signal
as the first input to the network. This dynamic spectrum is guaranteed
to contain a signal, because the minimum detection threshold
in
typical SETI 
searches is set at $\geq$ ten times the standard deviation of the
noise. Using
an estimate of the drift rate of the signal in this dynamic spectrum, we
extrapolate the expected detection frequency to the starting epoch of a different (typically
subsequent) scan. We then extract a portion of the
dynamic spectrum of the second scan
to use as the second input to our
network.
This portion has the same dimensions as those of the first input and is centered on the expected detection frequency.
The output of the network provides an assessment of
whether or not the second dynamic spectrum contains the same signal as
the first dynamic spectrum.
The CNN will be trained to perform this task with a large ($\sim$1 million) labeled
training set, such that the CNN can make this assessment by
recognizing patterns in the images, as opposed to calculating and
comparing estimates of signal properties 
like 
frequency and drift rate.

In what follows, we will use the terms 
``first'' or ``top'' image to refer to the dynamic spectrum of the
first scan, which must contain a signal of interest. Likewise, we will
use the terms ``second'' or ``bottom'' image to refer to the dynamic
spectrum of the second scan, which may or may not contain the same
signal.

\section{Data Preparation} \label{sec:data_prep}

\subsection{Observations} \label{subsec:obs}

We compiled our data set from the observations presented by
\citet{Pinchuk2019}. Those observations were conducted on 2017 May 4,
15:00 – 17:00 Universal time (UT) with the 100 m diameter Green Bank
Telescope. Both linear polarizations of the \textit{L}-band
receiver were recorded with the GUPPI back end in its baseband mode
\citep{GUPPI}. GUPPI was configured to channelize 800 MHz of recorded
bandwidth into 256 channels of 3.125 MHz each.

The observations primarily consisted of sources from the Kepler field, 
but also included scans of TRAPPIST-1 and
LHS 1140. A total of twelve 
sources were scanned. A full list of targets and their properties can be 
found in Table 1 of \citet{Pinchuk2019}.
This article also includes details relating to the formation of the
dynamic spectra, which have a time resolution of 0.336~s and frequency
resolution $\Delta \nu$ of 2.98~Hz.

A total of 10,293,618 signals were detected in the data, 
8,592,771 of which have a S/N $\geq 10$.
Because we had a large number of signals to choose from, we carefully
pruned the data according to principles described in Section~\ref{subsec:data_select} in order to obtain the best possible training candidates.

\subsection{Definition of Data Sets}\label{subsec:def_data_set}
In order to successfully train our CNN, we need to set aside several 
small portions of our data that we can use to evaluate the model performance
during and after training. Typically, it is recommended to set aside 
10--20$\%$ of the training data as a ``validation'' set that is used
to evaluate important metrics like precision, recall, and a model cost function or loss
\citep{HOML}.
These metrics can
then
be used to tune model hyperparameters (Section~\ref{subsec:hyperparam}) 
or identify problems like overfitting, which occurs when 
a neural network simply learns to reproduce the labels of the training data and 
therefore generalizes poorly to any other data. 
Standard ML practices
suggest that another 10--20$\%$ of the training data should be set aside 
as a ``test'' set that is only ever used to evaluate the performance of
the final model.
This evaluation is important in order to obtain an accurate estimate
of how well the model generalizes to data that it has never seen before.
 
When the training data are representative of the data that the model
will see in
a production environment, the training, validation, and test sets are
enough to successfully train a CNN from start to finish. 
However,
because
we needed to 
label
a large training set in an automatic manner (Section~\ref{subsubsec:image_pair_creation}),
our training data consist of two images taken from a single
scan, whereas the production data consist of two images from entirely
separate scans.
This difference between training and production data makes our application atypical and requires adjustments to standard ML practices.
In particular, 
it is possible for the network to
perform
well on the training data but poorly on the
production data. This situation occurs
when there is a ``data mismatch''
and often requires some manipulation
of the training set in order to better match the production
data. Importantly, this condition must be detected and addressed
before the model is put into production. If the validation set is
generated as a subset of
the training
set,
then data mismatch is
impossible to detect. On the other hand, if the validation data is
comprised only of data that the model will see in production,
it is not possible to discern whether poor model performance
is attributable to data mismatch or to model training issues, such as overfitting,
in the absence of other information.
The solution to this problem
is to create
an additional data set, the ``train--dev'' set,
which is a subset of 10--20$\%$ of the training data and is used to monitor the
performance of the model during training and detect problems like
overfitting.
With the ``train-dev'' set on hand, we can compile 
validation and test sets
that match the data that the model will see in production.
These two data sets are hand-labeled,
as is standard in most ML applications,
and are therefore much smaller than the
training and train--dev sets.
The validation and test sets
provide a useful
way to 
select an appropriate CNN architecture, 
tune hyperparameters, and
measure the model's generalization to new 
data,
among other uses.

We compiled a subset of signals from the hand-labeled validation and training sets to evaluate the performance of the UCLA SETI Group DoO filter.
In order to facilitate a
fair comparison to our CNN, we only kept the labeled image pairs that
corresponded to different scans of the same source, and we only
applied the persistence portion of the DoO filter to
these signals (the signal pairing logic is identical for both
components of the filter). This is important because the
DoO
filter
examines many different scans from a single
observing session to look for the presence of a given signal elsewhere
on the sky. However, when determining if the signal is persistent in
its detection direction, the filter only examines the two scans of the
source containing the signal of interest. The latter is consistent
with a standard application of the CNN trained in this work and
therefore offers the best comparison between the performance of the
existing DoO filter and the CNN.

We also made use of a small hand-labeled data set of image pairs to optimize 
and evaluate
a baseline model (Section~\ref{subsec:im_corr}). 
This model, which does not rely on ML techniques, provides a mechanism to test 
the performance improvement
due to the ML application.
Table~\ref{tab:data_sets} summarizes the data sets utilized in this work.

\begin{deluxetable}{lclc}[h!]
\caption{Data set name, size, and usage for all data sets presented in this work. 
Sections \ref{subsec:train_set_gen} and \ref{subsec:val_sets} detail the 
data compilation strategy, including the choice of data set size.
The parentheses in the ``Usage'' column specify a particular use case for the data set. The final column details whether or not the data set was labeled by hand. \added{All hand-labeled sets resemble the data processed by the model in a production setting.}
\label{tab:data_sets}}
\tablehead{
  \colhead{Name} & \colhead{Size} & \colhead{Usage} & \colhead{Hand-labeled? \deleted{Production-like?}}}
\startdata
Training             & 1,000,000 & Training and evaluating the CNN                       & No  \\
Train-dev            & 100,000   & Evaluating the CNN (Overfitting)                      & No  \\
Validation           & 1156     & Evaluating the CNN (Data mismatch)                    & Yes \\
Test                 & 1272     & Evaluating the CNN (Final results)                    & Yes \\
UCLA DoO test        & 1238     & Evaluating the UCLA direction-of-origin filter    & Yes \\
Baseline             & 524       & Optimizing and evaluating the baseline model          & Yes \\
\enddata
\end{deluxetable}
Note that the validation and test sets are much smaller than the training and 
train--dev sets because the former had to be analyzed and labeled by hand 
whereas the latter were labeled automatically, as described in 
Section~\ref{subsec:train_set_gen}.

\subsection{Data Selection Filters} \label{subsec:data_select} 
\begin{figure}[htb]
  \begin{center}
    \includegraphics[width=7in]{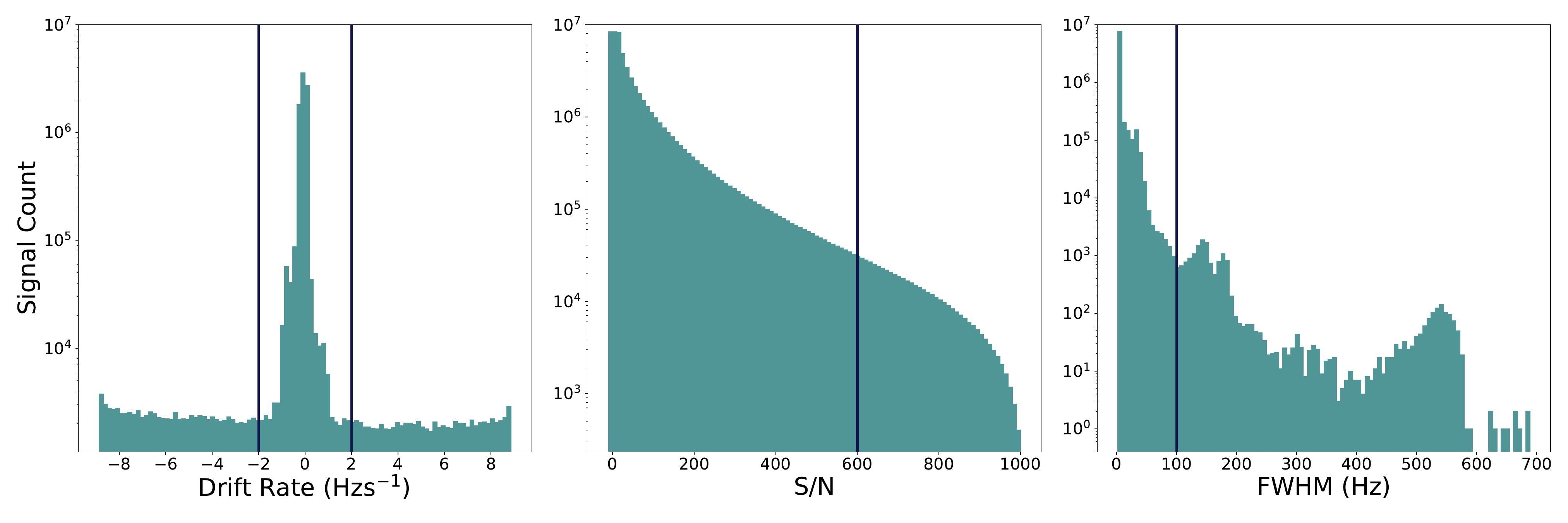}
\caption{Histograms of the (Left) drift rates, (Middle) S/N, and (Right) bandwidths as measured by a FWHM metric of detected signals.}
\label{fig:multi_hists}
\end{center}
  \end{figure}

We began by examining the distribution of drift rates of the
8,592,771 signals detected 
in 2017 
(Figure~\ref{fig:multi_hists}, Left).
We observed that the vast majority
($ > 98\%$) of detected signals have drift rates $| \dot f | \leq 2$ 
\Hzsns.
Moreover, by examining the
dynamic spectra of signals with drift
rates $| \dot f | > 2$ \Hzsns, we observed a lack of the narrowband 
characteristics that are
often chosen as
one possible
diagnostic of extraterrestrial engineered
emitters.
For these reasons, we excluded 168,684 ($1.96\%$) signals with drift
rates $| \dot f | > 2$ \Hzs from the training set.

We performed a similar
cut on the basis of the S/N of the detected signals. 
\replaced{Instead of significantly altering the S/N distribution, however,}{In particular,} 
we opted to remove any signals with \deleted{extremely} large S/N.
Based on the cumulative distribution of S/N shown in 
the middle panel of Figure~\ref{fig:multi_hists}
 we chose to discard any signals with a S/N $> 600$. 
This threshold was large enough to preserve $> 99\%$ of the remaining 
signals but also remove any extreme S/N outliers.
As a result of this filter, we removed 58,399 ($0.69\%$) signals with extreme S/N from 
the training set.

Next, we examined the bandwidth of the remaining signals, as
quantified by a full width at half maximum (FWHM) metric.  The
bandwidth of the training signals is especially important because we
needed our training-set signals to fit within a 225$\times$225 image. 
This image size was chosen to satisfy a few considerations.  
First, the number of \replaced{lines}{rows} cannot exceed half the number of 
\replaced{lines}{rows} in the 
dynamic spectra, which is $\sim$490 for the 2017 data at 2.98 Hz frequency 
resolution. Second, the number of columns must be an odd number so that signals 
can be perfectly centered in the image. Third, the input size for models like 
ResNet52 are images of size 224$\times$224 pixels \citep{resnet} and have been 
shown to be manageable with modern CNN architectures.
Our chosen image size corresponds to an upper limit of $\sim$670 Hz on the
bandwidth of the signals.
However, because we are only interested in
narrowband signals
($\leq10$~Hz) in this work,
we can set the bandwidth threshold much
lower than the theoretical upper limit. 
The distribution shown in the right panel of Figure~\ref{fig:multi_hists}
suggests a threshold value of 100 Hz. As a result of this filter, we
removed 15,044 ($0.18\%$) signals
with large bandwidths from the training set.

Finally, we discarded 6447 signals that were detected close enough to the 
edge of their 3.125 MHz wide channel such that the signal
overlapped with a
neighboring channel. Although we could ``stitch'' neighboring channels
together to fully recover these signals, given the vast number of signals
left to choose from, we decided to simply remove this tiny fraction 
of signals from consideration
for the purpose of building the training set.
When we evaluate actual data with the CNN, we do combine two channels
when necessary to correctly represent signals located near the channel
edges.

Overall, the filters above discarded 248,594 of the 8,592,771 detected 
signals, leaving 8,344,177 ($97.11\%$) available to use for our training 
set. If we assume that the distributions of drift rates, S/N, and bandwidths
remain relatively constant
over time
and the RFI observed with the 2017 antenna pointing directions is representative of other directions,
then the network trained on our pruned data set
should be applicable to $\sim$97\% of the detected signals in future
searches with similar parameters.  If the RFI environment changes so
much in time or in space that it severely alters the distributions of
signal properties, this percentage value could change.
Although the RFI environment
may evolve
in
time
and
space, we were able to verify that these filters still
captured $\sim$97\% of the detected signals in searches conducted in
2018 and 2019 by \citet{Margot2021}
with different antenna pointing directions.
Specifically, 97.4\% (9,845,561 out of 10,113,551) and 97.2\%
(16,048,515 out of 16,518,362) of the signals detected in
2018 and 2019, respectively, passed the data selection filters
described above.

The range of signals selected for the labeled training set translates
into a finite domain of applicability for the CNN. Because the CNN was
not designed for signals with S/N $<$ 10, frequency drift rates $| \dot
f | > 2$ \Hzs, or bandwidths $>$ 100 Hz, it may not perform well when
applied to such signals.
\added{Therefore, we recommend eliminating signals that are outside the domain of applicability of the CNN prior to applying the CNN.}
Note that the applicability of the training set to new data sets and
the CNN's decision accuracy are two different concepts.  We evaluate
the latter in Section~\ref{subsec:model_perf}.

\subsection{Generation of the Training and Train--dev Sets} \label{subsec:train_set_gen}
Labeling a sizable training set by hand is
time-consuming.
To bypass this limitation, we developed a strategy to synthetically 
generate our labeled data set. 
Because we are interested in training a neural network to
supplement our DoO filters by detecting whether or a not a 
signal is present in two separate images, we need a training set that
consists of pairs of images labeled with a binary flag indicating
the persistence of a signal across both scans.

\subsubsection{Creation of Image Pairs}\label{subsubsec:image_pair_creation}

In order to simulate a pair of scans containing the same signal, we
split the image representing a single scan into two 
parts 
along the
time dimension.  We then evaluated whether the signal was detected in
both the top and bottom parts. If the signal was detected in both,
we labeled the pair as a positive sample. Otherwise, we
labeled the pair as a negative sample to signify that the
signal was present in only one of the two parts.  In practice, the
detection decisions are implemented by computing the ratio of signal
powers in the top and bottom parts.

The power ratio calculations rely on the simplifying assumption that the 
total integrated power
associated with each signal is distributed evenly 
throughout the duration of the scan. In other words, we expect the signal in
each half of the spectrum to contribute equally to the total power.
We
calculated the 
signal power detected in each half
of the scan and recorded
these values as power ratios ($P_{\rm top}, \, P_{\rm bottom}$),
where the denominator is half of the total signal power in the scan.
The signal power was calculated by summing $2n+1$ pixels at each
timestep along a line with a slope equal to the drift rate of the
signal, where $n = \left \lfloor 1.5 \times \frac{B_{\rm
    FWHM}}{2\Delta \nu} \right \rfloor$ pixels on either side of the
line,
where $B_{\rm FWHM}$ is the bandwidth of the signal measured
in hertz, $\Delta \nu$ = 2.98~Hz is the frequency resolution of the
data,
and $\lfloor \ \rfloor$ is the floor operator.
The drift rate of the signal was assumed to be the same in both
halves of the spectrum.  To calculate the signal power in the the top
half of the dynamic spectrum, we started at the pixel corresponding to
the detection frequency of the signal.  To calculate the power of the
signal in the bottom half, we started 
at the center frequency obtained by
linearly extrapolating the signal detection
frequency
to the appropriate time.  The total signal
power was obtained by summing the two halves.
By comparing the ratio of powers in the top and bottom halves of
each scan to a suitable threshold, we were able to assign an appropriate label to each signal.
Section~\ref{subsubsec:signal_selection} describes the selection of the threshold.

This approach allowed us to label a large number of signals in a short 
period of time.
We chose to compile the 
training and train--dev
data sets
from a pool of
1,100,000 total signals,
which is a random selection among the 8,344,177 signals that meet certain 
criteria described below.
For reference, the popular MNIST handwritten digit dataset
\citep{MNIST} as well as the CIFAR--10 \citep{CIFAR10} 
multiclass image dataset both contain 60,000 
samples each.
Both the training and train--dev sets contain an equal ratio of
positive and negative samples. We set aside 100,000 signals 
for the train--dev set, leaving 1 million signals to be used as the training 
set.

In order to be accepted into the
training set or train--dev
data set,
signals had to satisfy several criteria.  Most importantly, the top
image
in a pair, which mimics the first scan in an actual observing
sequence, must always contain a signal.  Moreover,
the primary signal must always be centered in the top image and nearly
centered in the bottom image, i.e., the signal must start in or near
the middle of the frequency array in the topmost time bin.  This
requirement affects the construction and processing of the images, which are 
described in Section~\ref{subsubsec:proc_selected_signals}.
In particular, we allowed a small tolerance on the location of the
signal in the bottom image, but the signal in the top image must
always start at column 113 (if counting from 1) in the first row of the 
225$\times$225 images.
Both of these criteria can easily be met in
production, because
one can apply
these cropping steps
to detected signals
with known starting frequencies.
For signals whose bandwidth spans several pixels, the starting
frequency is defined as the starting frequency reported by the
detection algorithm, which is where most of the power is detected 
\citep{Margot2021}.

\subsubsection{Selection of Suitable Signals}\label{subsubsec:signal_selection}
\begin{figure}[htb]
  \begin{center}
    \includegraphics[width=3in]{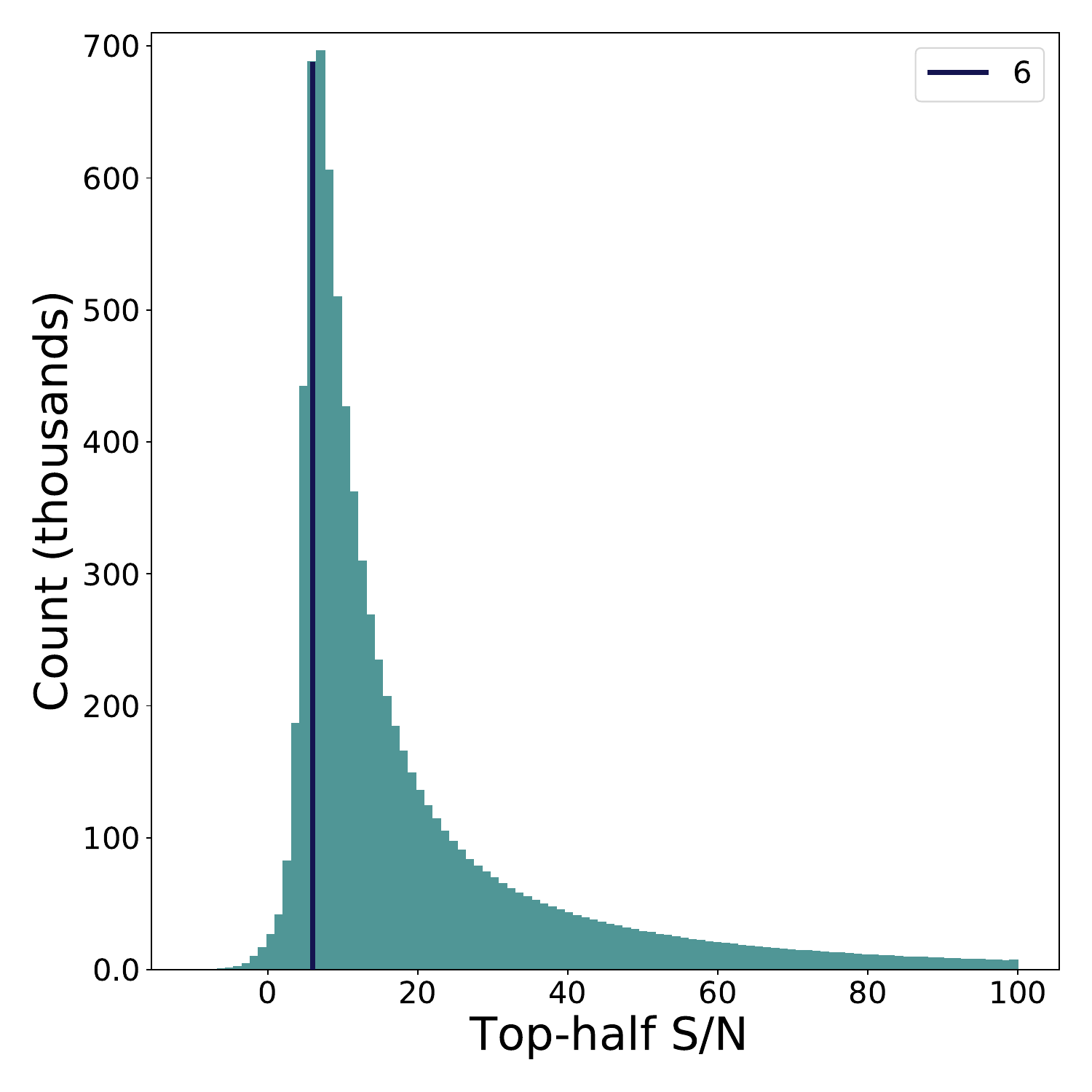}
\caption{Histogram of the S/N detected in the top half of each scan. The blue vertical line shows the S/N cutoff value of 6 used to remove signals with low power in the top half of the scan.}
\label{fig:top_half_snr_hist}
\end{center}
  \end{figure}

We began by examining the S/N of each signal in
the top half of the scan only.
In order to satisfy the underlying assumption
that there is definitely a signal in the center of the first image 
(represented in the training set by the top half of the scan), 
we required a minimum top-half
S/N of
at least 6
(Figure~\ref{fig:top_half_snr_hist}), which corresponds to a $\sim$1 in a billion 
false detection rate.
We validated our choice of threshold by examining a sample of 
signals below the cutoff value. We found that most of these signals are faint and 
difficult to
detect visually,
while the rest are not present at all.  On the contrary, signals
above this threshold are clearly visible in the
dynamic spectra.  Appendix~\ref{app:example_sigs} illustrates these two cases.

\begin{figure}[htb]
  \begin{center}
    \includegraphics[width=3in]{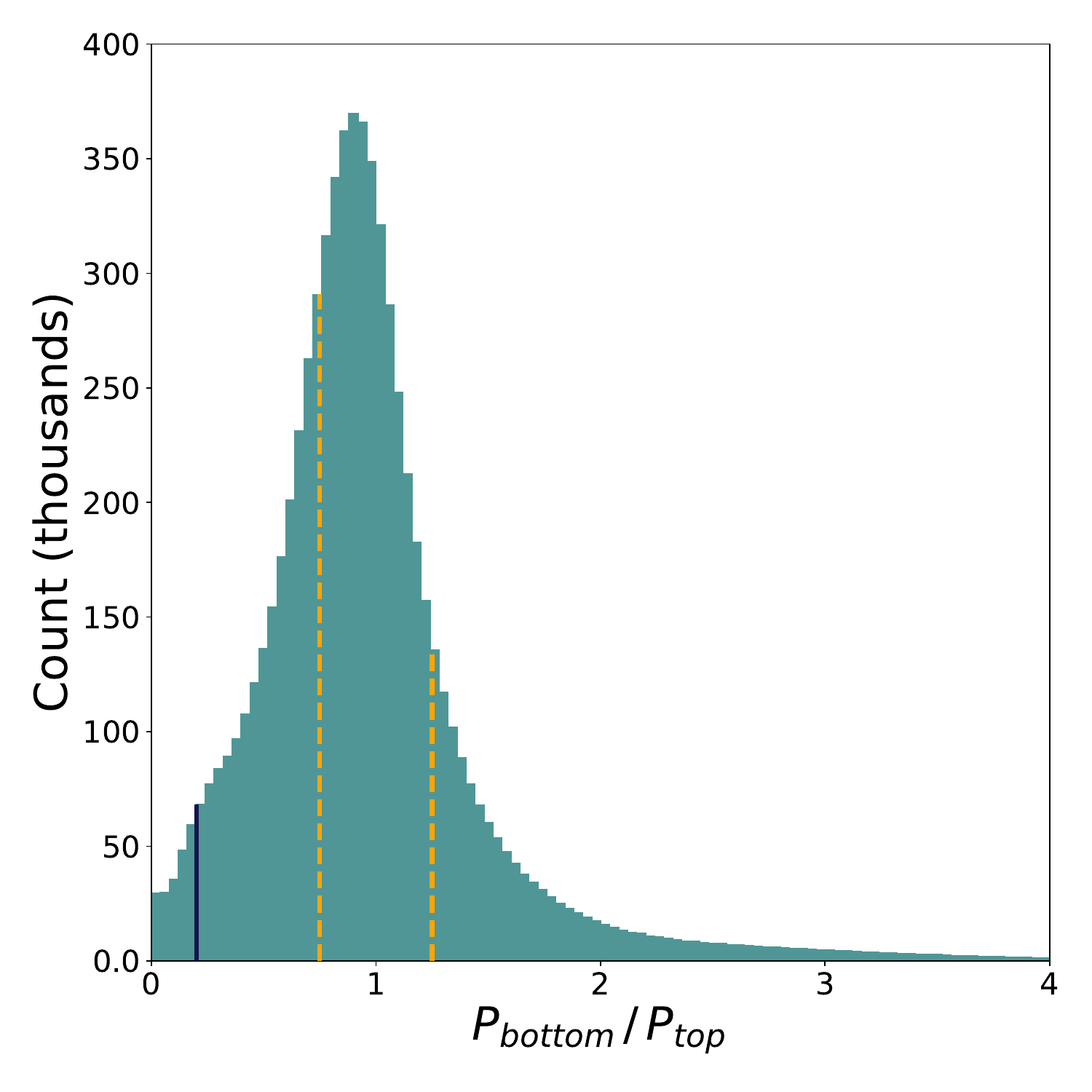}
    \caption{Distribution of the ratio of integrated powers, or $P_{\rm bottom} / P_{\rm top}$ ratio. The \added{dashed} orange vertical lines delimit the lower and upper bounds that we used to select 500,000 positive samples (0.75 and 1.25, respectively).  
The \added{solid} blue vertical line is plotted at a ratio of 0.2. We selected 50,000 
signals below this value to represent a portion of our negative samples.}
\label{fig:bot_top_dist}
\end{center}
  \end{figure}

As we performed our final selection, we needed to allow for variations
in the S/N of the top and bottom portions of the positive signals,
since the top and bottom portions represent two separate scans and
we have empirically observed that the S/N can change substantially between 
scans. To do so, we compared the integrated power values from the 
top and bottom halves of each signal. Specifically, we examined the 
distribution of the ratio of bottom to top
integrated powers (i.e., $P_{\rm bottom} / P_{\rm top}$).
We found that
approximately 50\% of the signals have a ratio between 0.75 and 1.25
(Figure~\ref{fig:bot_top_dist}), so we randomly
selected 550,000
signals from this region to represent our positive samples (i.e., a
signal is detected in both scans/images).
Additionally, we selected 50,000 signals with a ratio of 0.2 or lower to
partially represent our set of negative samples 
(see blue line in Figure~\ref{fig:bot_top_dist}). 
To ensure the 
absence of a signal from the
bottom half,
we verified that none
of these samples had any signals with a prominence value greater than 
3 times the standard deviation of the noise
in the bottom half (see \citet{Margot2021}, Section 3.1
for an in-depth discussion of the prominence calculation).
Appendix~\ref{app:example_sigs} depicts a sample of these signals.
Ideally, \textit{all} negative samples would be
obtained with this method, but there were not enough of these signals
to provide the necessary negative samples.
Since this category is grossly underrepresented in the data,
we used data augmentation to create more negative samples. Specifically,
the remaining 500,000 negative samples were obtained by taking samples
from the region with a $P_{\rm bottom} / P_{\rm top}$ ratio between 0.75 and 
1.25 and altering them in four different ways to remove any signals present in 
the bottom half. This process is described below.

\subsubsection{Processing of Selected Signals}\label{subsubsec:proc_selected_signals}
After selecting the signals for our training set, we applied some
processing to ensure that signals in the positive and negative
categories were representative of their respective labels.

We first considered the 550,000 signals from the positive category.
When we applied our ML algorithm to real data, we
obtained the bottom image by extracting a portion of the spectrum from
the second scan centered on the frequency value calculated by
extrapolating the frequency detected in the first scan.
If the
same signal is present in both scans,
and if the signal's drift in time-frequency space is approximately
linear\footnote{The assumption of linearity is reasonable for a source analyzed at \textit{L} band with a frequency resolution of $\sim$2.98 Hz, a scan duration of 150~s, and a line-of-sight jerk below $2.311 e^{-5} \rm ms^{-3}$.
For reference, the maximum line-of-sight jerks for Earth's spin and orbit are $2.453 e^{-6} \rm ms^{-3}$ and $1.181 e^{-9} \rm ms^{-3}$, respectively.}, 
and if the drift rate estimate is approximately correct,
this method will ensure that the signal also appears in the bottom image,
but it does not guarantee that the signal will be
perfectly \textit{centered} in
the bottom image.
For instance, a small discrepancy between the actual and estimated
drift rates can result in an offset between predicted and actual
frequency values.  To simulate this scenario in our training set, we
shifted all of the signals in the bottom image of our samples by
$\pm$0--5 pixels (0--15 Hz). The exact shift for each image was
randomly selected
from a distribution of shift values 
described in Section~\ref{subsec:im_corr}.

The 550,000 signals for the negative category were compiled using five distinct 
procedures. The first 50,000 signals were selected from the distribution 
shown in Figure~\ref{fig:bot_top_dist} with a $P_{\rm bottom} / P_{\rm top}$ 
ratio of $<0.2$. For each of these signals, we verified the lack of any signals 
with a prominence value greater than
3 times the standard deviation of the noise
in the bottom half of the 
scan. The next 125,000 negative samples were obtained by selecting unused 
signals from the ``positive'' range 
($0.75 \leq P_{\rm bottom} / P_{\rm top} \leq 1.25$) and shifting the signal in 
the bottom image by $\pm$6--10 pixels (18--30 Hz).
By doing so, we 
forced the algorithm to learn that
a positive detection requires the bottom signal
to be detected in
close proximity to the
extrapolated frequency,
which is calculated on the basis of signal properties in the top image.
We obtained another 125,000 negative samples by once again selecting 
unused signals from the ``positive'' range and replacing the bottom signal 
with 
an
unrelated signal (also sampled from the ``positive'' range). 
This group of negative samples
forced the algorithm to compare signal
properties and \textit{not} pair two unrelated signals that may have been 
detected at similar frequencies in two different scans. Another 125,000 
negative samples were obtained by selecting leftover signals from the 
``positive'' range and replacing the bottom image
with noise.  The noise was generated by sampling values from a
$\chi^2$ distribution
with four degrees of freedom 
that was fit to the bottom image after removing
any power values belonging to any signals detected in the
spectrum. 
The signals were removed by obtaining the database records of all
signals detected within the relevant portion of the spectrum 
and discarding any power values within 2 times the measured
bandwidth along the linear drift rate of each signal.
The final 125,000 samples were obtained similarly, but
instead of replacing the entire bottom image with noise, only the
power values belonging to the signal in the bottom image were replaced
with sample values from 
a $\chi^2$ distribution that was fit to the bottom image with the same
procedure
as above.

\begin{figure}[htb]
\begin{center}
  \includegraphics[width=7in]{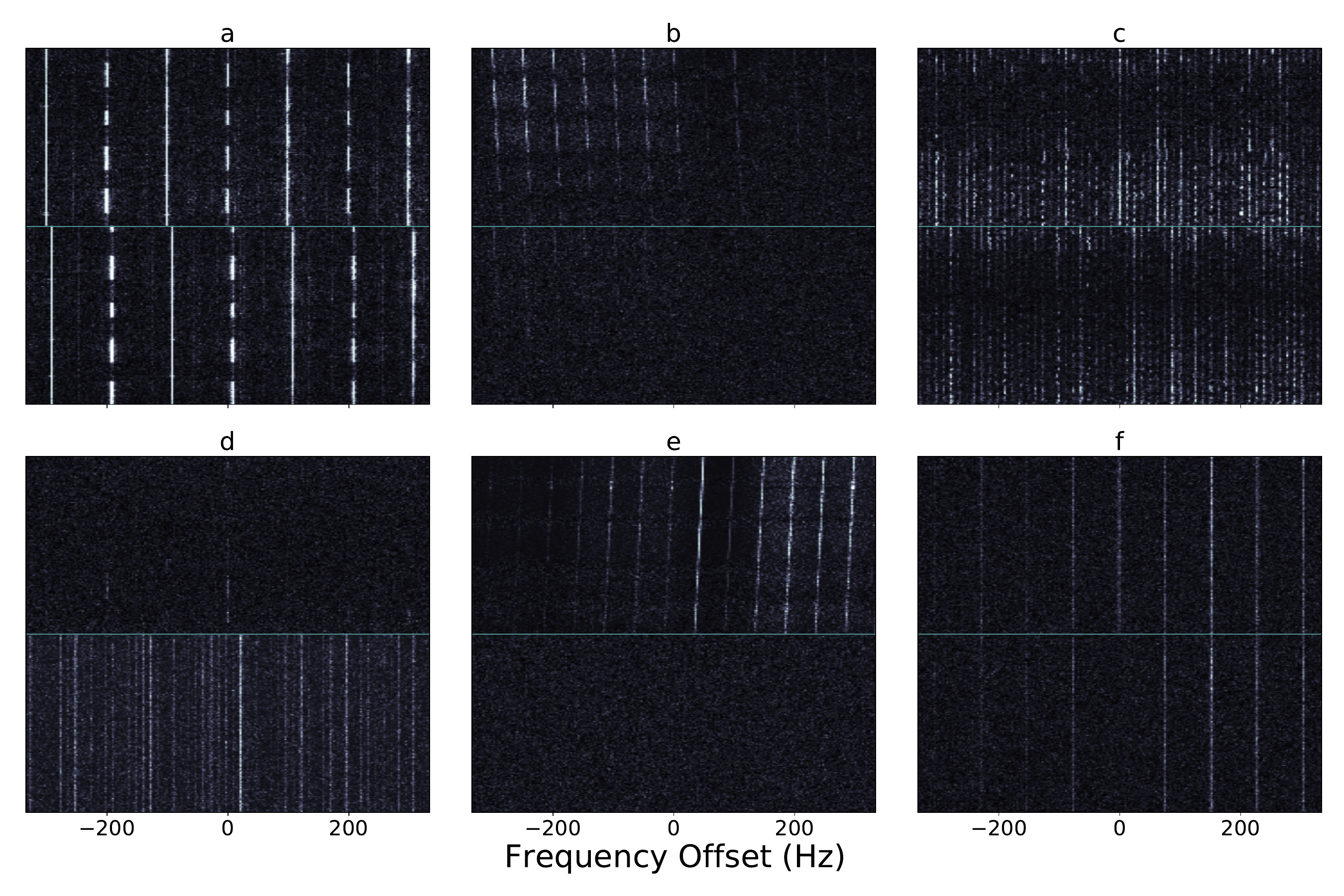}
\caption{
Sample signals used in the ML labeled training set.
(a) Signal from the positive category, shifted 3 pixels ($\sim$9 Hz) to the right. (b) Sample signal from the negative category with a $P_{\rm bottom} / P_{\rm top}$ ratio
  of $<0.2$. (c) Sample signal from the negative category, shifted 8 pixels ($\sim$24 Hz) to the right. (d) Sample signal from the negative category with an unrelated signal in the bottom image. (e) Sample signal from the negative category with a bottom image consisting completely of
  simulated noise. (g) Sample signal from the negative category with the primary (center) signal replaced by noise in the bottom image.}
\label{fig:data_proc_examples}
\end{center}
\end{figure}

An example product of each of the above procedures is shown in 
Figure~\ref{fig:data_proc_examples}.

\begin{figure}[htb]
  \begin{center}
    \includegraphics[width=7in]{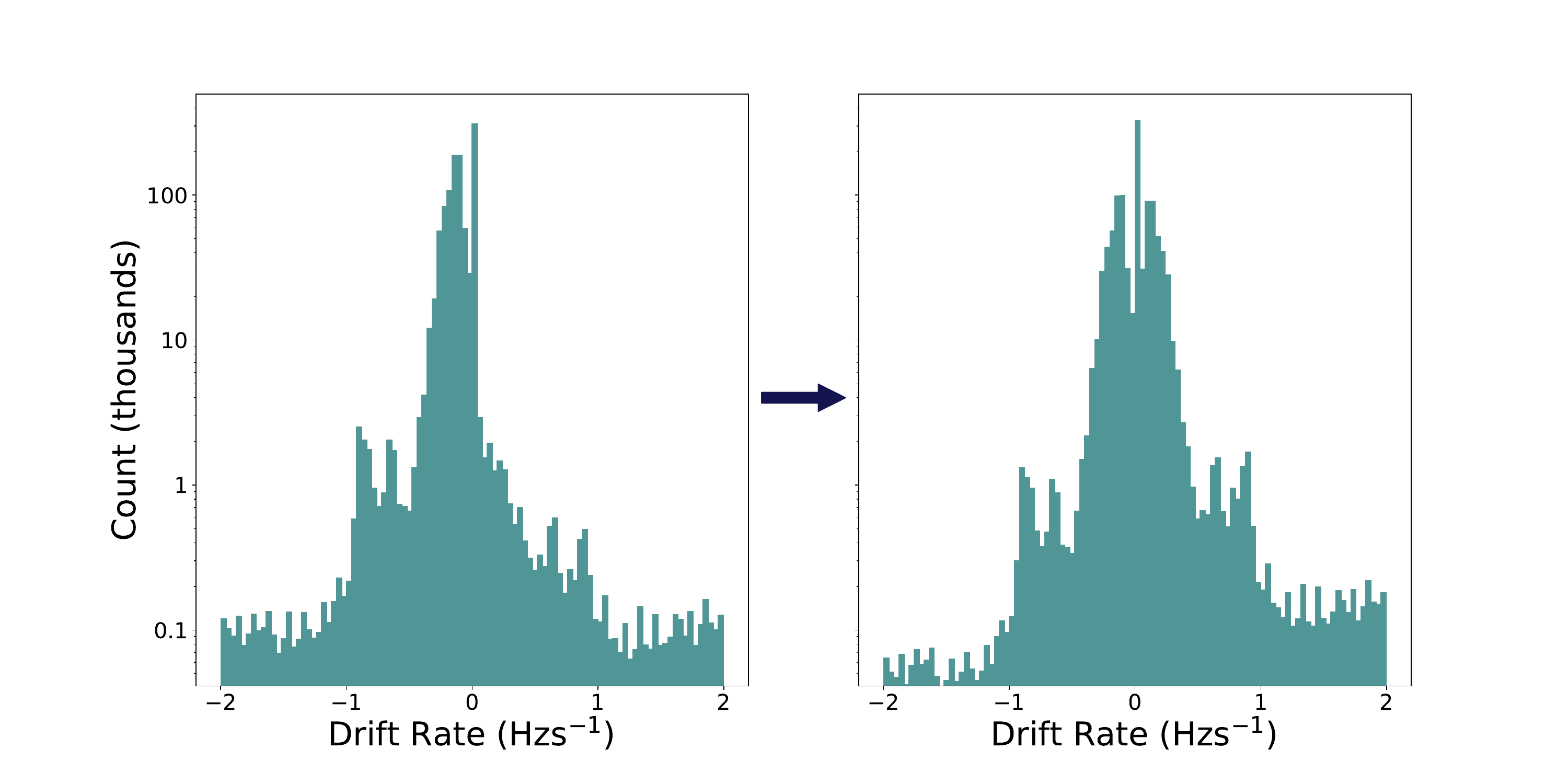}
\caption{
(Left) Drift rate distribution of the 1.1 million signals selected to be part of our training set. Note a significant bias toward signals with a negative drift rate value. (Right) Drift rate distribution of the same set of signals after applying a horizontal flip to 364,184 negative drift rate signals. The bias 
that affects 
$\sim$685,000 signals
between $-0.5$ and $0$ \Hzs
is almost entirely removed
at the expense of a slight bias introduced between $-2$ and $-1$ \Hzsns
that affects 
$\sim$2500 signals.
}
\label{fig:dfdt_bias}
\end{center}
  \end{figure}

Before finalizing the training and train--dev sets, we examined the drift rate
distribution of the 1.1 million signals selected 
with the process described above.
This distribution is biased
toward signals with negative drift rates (Figure~\ref{fig:dfdt_bias}; left).
This bias is expected from most low- and medium-Earth-orbit satellites, such as global positioning system satellites, 
which orbit in a prograde fashion with respect to the telescope.
In order
to avoid inadvertently introducing this bias into our model, we
selected 364,184 signals with a negative drift rate using
a stratified split \citep{HOML} on the signal drift rates,
and horizontally flipped the images corresponding to these signals. The 
resulting drift rate distribution exhibited a significantly reduced bias
between $-0.5$ and $0$ \Hzs at the expense of a slight bias between $-2$ and 
$-1$ \Hzs (Figure~\ref{fig:dfdt_bias}; right).

At the end of the compilation process, our training set consisted of 550,000 
positive samples and 550,000 negative samples. The 1.1 million samples were
separated into 1 million training samples and 100,000 train--dev samples.
The samples were separated using a stratified split on the bandwidth of the 
signals. Each set contained an equal number of positive and negative samples.

\subsection{Creation of Validation, Test, and Baseline Model Data Sets}\label{subsec:val_sets}
 
We compiled a small set
of 1156 hand-labeled images, where the top and bottom images are extracted
from two separate scans. These images are a true representation of the
samples that the network will see during production, so we use them as
our validation set.
By comparing the model performance on the
train--dev and the hand-labeled validation set, we can
assess
whether or not there is a mismatch between the training set and the real test 
data. 
We also compiled a set of 1272 hand-labeled images to serve
as the test set. These samples contain signals from two different scans, as 
would be the case in the production environment. 
Finally, we selected 524 hand-labeled samples,
where each sample contains a signal 
in both scans,
to optimize and evaluate our baseline model 
(Section~\ref{subsec:im_corr}). This data set has no samples in common with any of the 
4 data sets described above.

\section{Models} \label{sec:models}

\subsection{Baseline Model} \label{subsec:im_corr}
We devised a simple correlation-based model to serve as a
benchmark or 
baseline for our
results.
We began by selecting 
a baseline data set as described in Section~\ref{subsec:val_sets}.
We then calculated the 2D correlation coefficient between the
two signals in each data sample, using
all available time steps in 
a region of 
frequency
width
$2w + 1$ centered on each signal. The 
correlation coefficient is given by
\begin{equation}\label{eq:corr_coeff}
\rho(A, B) = \frac{1}{N - 1}\sum_{i=1}^{N}\left(\frac{A_i - \mu_A}{\sigma_A}\right)\left(\frac{B_i - \mu_B}{\sigma_B}\right),
\end{equation}
where $A_i$ and $B_i$ are the individual pixels of images $A$ and $B$,
$\mu_j$ and $\sigma_j$ 
are the mean and standard deviation of the pixels
under consideration in
image 
$j$, $j \in \{A, B\}$,
respectively, and $N$ is the total number of 
pixels compared.
To ensure that our results were not influenced by poor 
localization of the signals in the images, we shifted the bottom image by 
$\pm 3$ pixels
in the frequency dimension
and computed $\rho(A, B)$ in each case. We report the maximum 
correlation score from the set of seven resulting values. We tested both a 
large
($w=15$ pixels, $\sim$50 Hz) and small ($w=3$ pixels, $\sim$10 Hz) window size, and found that
the latter gave the best results
in terms of model 
precision and recall.

After computing the correlation values, we selected a threshold value 
in order to assign a label for each set of images. The label is positive 
(i.e., ``True'', 1) if the signals in the images are strongly correlated,
or negative (i.e., ``False'', 0) if the signals in the images 
are unrelated. Typically, this threshold is chosen by finding 
the best trade-off between precision and recall
\citep{HOML}.
Precision is defined as the ratio of the true-positive count
(i.e., label=prediction=1) to the sum of the true-positive and false-positive counts (i.e., prediction=1). In other words, when a model with
a precision value of 1 predicts that an image pair belongs to the
positive class, it is always correct. On the other hand, recall is
defined as the ratio of the true-positive count to the sum of the true-positive and false-negative counts (i.e., label=1). A model with a
recall value of 1 will always correctly
classify
all the positive
samples.  A perfect model would have both recall and precision values
of 1. In practice, there is always a trade-off between the two
metrics.

\begin{figure}[htb]
\begin{center}
    \includegraphics[width=5in]{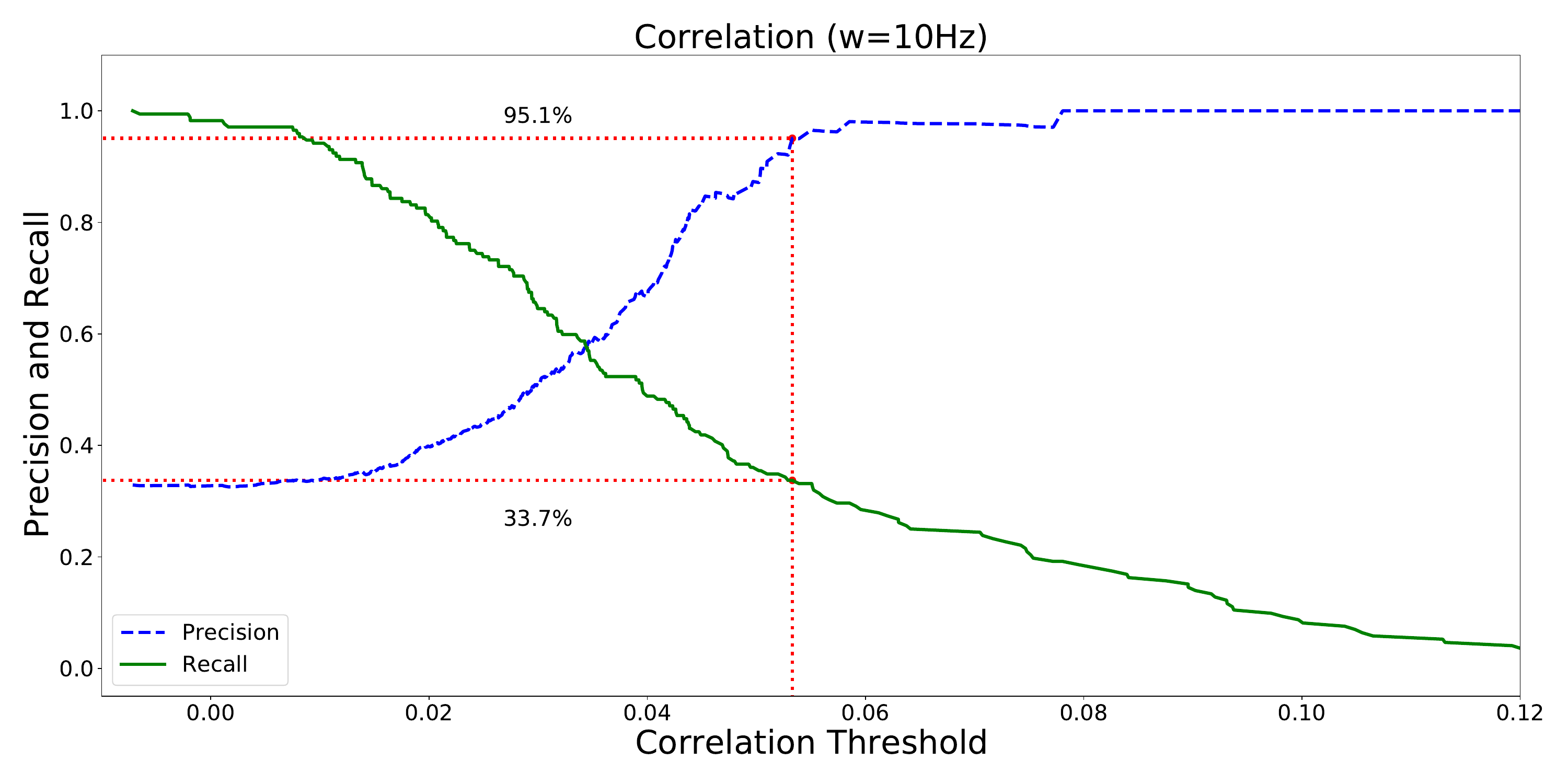}
    \caption{Precision and recall curves for
      a baseline 2D correlation model,
      which does not rely on ML techniques and is used solely to serve as a benchmark 
to evaluate the performance improvement of our ML application (Section~\ref{subsec:model_perf}).
With the chosen threshold, the baseline model detects approximately a third of 
the RFI in the data with 95\% precision.}
\label{fig:pr_baseline}
\end{center}
\end{figure}

In our application, precision is more important 
than recall because a larger precision value minimizes the number of 
false positives. False positives represent valid candidate technosignature 
signals that were only detected in one image (or direction of the sky), 
yet were still
classified
as RFI. For this 
reason we chose our threshold as the correlation value that yielded a precision 
$\geq 95\%$.
At this threshold (0.0551), the recall was 33.7\% 
(Figure~\ref{fig:pr_baseline}). In other words,
the baseline model only detects 
$\sim$1/3 of the RFI in the data, but it does so with 95\% precision.

The baseline model also helped define a distribution of frequency
shifts that we used in building the training and train--dev sets
(Section~\ref{subsubsec:proc_selected_signals}).  We selected a subset
of 5750 signals from the 2017 observations \citep{Pinchuk2019} that
passed the UCLA DoO filter and had correlation values that exceeded
the threshold of 0.0551.  We used randomly selected values from the
distribution of shifts of this subset to shift the signals in the
second images of our training and train--dev sets
(Section~\ref{subsubsec:proc_selected_signals}).

\subsection{Model Selection} \label{subsec:model_selection}

In order to select the best suitable model for the
DoO filter, we carried out a scaled-down performance
comparison of over 20 model architectures. For this comparison, we
selected four ResNet variants \citep[ResNet34, ResNet50, ResNet101,
  and ResNet152;][]{resnet}, two VGG variants \citep[VGG16,
  VGG19;][]{vgg}, and the Xception architecture \citep{xception}.  In
addition, we trained two Siamese model variants 
(see Appendix~\ref{app:siamese_models}) for each of these 7 models.  We did not
perform any hyperparameter tuning at this stage,
and we trained each model on only 10\% of the training set,
using 10\% of the train--dev set and the full validation set 
to evaluate the results,
because the goal was to quickly compare as many models as possible.

We found that all models outperformed their respective Siamese versions
when comparing the loss and
other relevant metrics.
We also found that none of the seven 
standard model architectures significantly outperformed the others in terms of these
metrics. However, we did notice that the Xception architecture did not exhibit significant 
overfitting during training, whereas all other models did.
Although there are multiple model 
regularization techniques designed to overcome model overfitting, we decided to select
the Xception model as our base architecture in order to reduce the amount of
model tuning required later during training.

\subsection{Hyperparameter Tuning} \label{subsec:hyperparam}

After selecting the best model architecture for the
DoO filter,
we were left with a significant number of hyperparameters to tune.
\citet{HOML} defined a hyperparameter as ``a parameter of a learning 
algorithm (not of the model). As such, it is not affected by the learning 
algorithm itself; it must be set prior to training and remains constant 
during training.''
All of the hyperparameters that were considered during this process, as well 
as several suitable values for each, are listed in Table~\ref{tab:hyperparams}.

\begin{deluxetable}{lcc}[h!]
\caption{Hyperparameters that were considered in this work, as well as the
set of possible values for each and the final value used for training the
model.
For a definition of these concepts, see \citet{HOML}.
\label{tab:hyperparams}}
\tablehead{
\colhead{Hyperparameter} & \colhead{Possible values} & \colhead{Final Value}}
\startdata
Optimizer                             &  \{Stochastic Gradient Descent, RMSProp, Adam, Nadam, AdaMax\}  &  Nadam      \\
Learning rate                         &  \{$1e^{-3}, 1e^{-4}, 1e^{-5}, 1e^{-6}$\}                       &  $1e^{-3}$  \\
Batch size                            &  \{16, 32, 64, 128\}                                            &  16         \\
Activation function                   &  \{ReLu, Swish\}                                                &  ReLu       \\
Fully connected layers on top         &  \{{\em True}, {\em False}\}                                            &  {\em False}    \\
Dropout rate                          &  \{{\em None},  0.2, 0.5\}                                          &  0.2        \\
Include squeeze-and-excitation blocks &  \{{\em True}, {\em False}\}                                            &  {\em True}     \\
Include input batch normalization     &  \{{\em True}, {\em False}\}                                            &  {\em True}     \\
\enddata
\end{deluxetable}

The ``Optimizer'', ``Learning rate'', ``Batch size'', and ``Activation 
function'' hyperparameters simply refer to the network hyperparameter that was 
tuned during this process.
The hyperparameter ``Fully connected layers on top'' refers to the addition of
one or more fully connected layers of neurons inserted immediately after the
global average pooling layer but before the final prediction node. The number
of layers and the number of neurons per layer were also tuned as part
of this process. 
When the hyperparameter ``Dropout rate'' was set to {\em None}, no changes were
made to the network architecture. Otherwise, a dropout layer \citep{dropout} 
was added at the end of the network with the corresponding dropout rate.  
The hyperparameter ``Include squeeze-and-excitation blocks'' refers to the 
addition of squeeze-and-excitation (SE) blocks \citep{sneHu18} at the end of 
every separable convolution\footnote{A convolution layer is the central building block of a CNN. It applies a convolution kernel to each pixel of an input image and produces a feature map.  When the input image contains multiple channels (e.g., red, green, blue),
the convolution kernel has a third dimension equal to the number of channels.}
module of the Xception architecture.  
The SE blocks are network units that are designed to adaptively recalibrate 
channel-wise feature responses by explicitly modeling interdependencies between 
the channels. \citet{sneHu18} demonstrated that SE 
blocks bring significant improvements in performance for state-of-the-art CNNs 
with only a slight addition to the computational cost. 
The ``input batch normalization'' hyperparameter controlled the 
normalization of the input data. Specifically, if this parameter was set to 
{\em False}, the input data would be normalized to zero mean and unit standard 
deviation, and no further modifications were made to the base network 
structure. When this parameter was set to {\em True}, the input data
were not 
scaled, but an extra batch normalization layer \citep{bn} was added immediately after the input layer of the network.

While a comprehensive grid search for the best hyperparameter combination would
yield the
optimal model configuration, we found that hardware limitations 
made this approach impractical.
A single training session with only 20\% of the training data and 10 epochs,
where each epoch represents a full pass of the training data
through the neural network,
took $\sim$10 hours
on a single
ML-enabled graphical processing unit (GeForce RTX 2060 SUPER 8 GB GPU),
which
would make a grid search 
prohibitively
large
considering the need to examine $\sim$4000 combinations.
With the current specifications, a grid search for the best hyperparameter 
combination from 
the set of values described in Table~\ref{tab:hyperparams} 
would take $\sim$4.5 years to complete.
Instead, we chose the 
best hyperparameter combination from the results of $\sim$30 different 
training sessions
of 10--15 epochs each
using judiciously chosen combinations of hyperparameters.
Our selection approach was ``semi-greedy'' because we allowed the results of 
previous training sessions to have some influence over the hyperparameter 
choice for the next session. 
\added{In general, we selected the hyperparameter value with the highest classification score for subsequent trials, but we also occasionally expanded the search space to include suboptimal values.}
\replaced{Although}{We emphasize that} this approach \added{includes some degree of arbitrariness and} does not guarantee a globally optimal model 
configuration\replaced{,}{. Still,} we found that the hyperparameter combination obtained via 
this method yields satisfactory model performance 
(see Section~\ref{subsec:model_perf}).

The final combination of hyperparameters was 
determined by comparing the model performance over all
$\sim$30
training sessions.
The best values for each parameters are listed in the final column of 
Table~\ref{tab:hyperparams}.

\subsection{Final Model} \label{subsec:final_model}
\begin{figure}[hbt]
\begin{center}
    \includegraphics[width=7in]{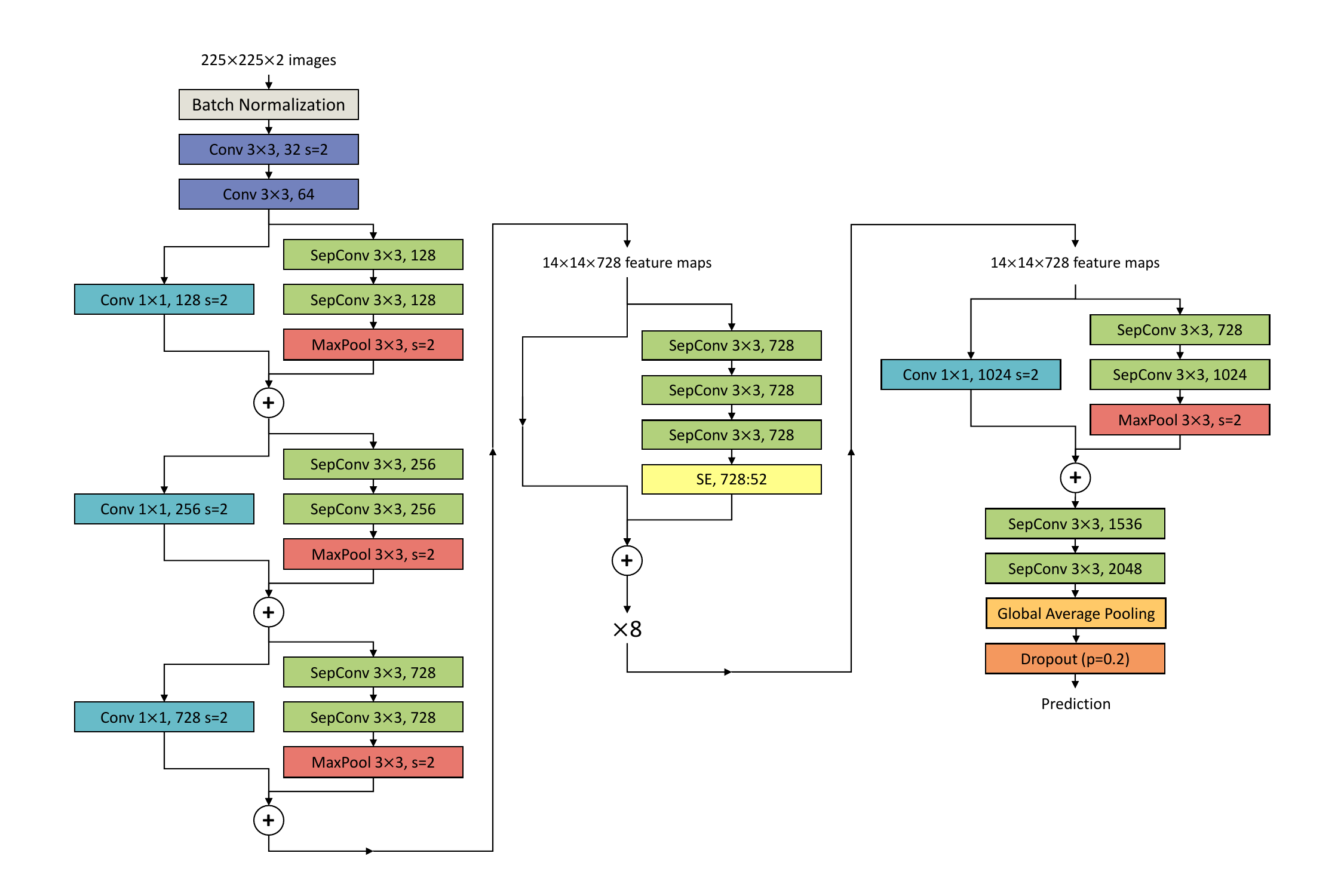}
    \caption{Architecture of the final model presented in this work. This figure is adapted from Figure 5 of \citet{xception}. Batch normalization and activation (ReLu) layers that follow each convolution and separable convolution layer are omitted from the diagram. Data flow follows the arrows. The middle portion of the network is repeated 
to create 8 identical sections.
For each layer, we list the name, the kernel size, and the number of output channels. Layers with a stride length of 2 instead of 1 are distinguished by ``s=2.'' The reduction ratio (14) of the SE layer is presented as ``728:52'', which denotes the number of input and output channels as a ratio of the number of hidden layer channels \citep{sneHu18}.}
\label{fig:model_architecture}
\end{center}
\end{figure}

Our final model architecture is shown in Figure~\ref{fig:model_architecture}.
The most important layer of the Xception architecture is the separable
convolution layer,
which consists of a spatial convolution performed 
independently over
each channel of an input,
where a channel refers to a slice along the depth dimension of the input
matrix, 
followed by a
1$\times$1 convolution 
projecting the 
outputs of
the first convolution onto a new 
space. \citet{xception} argued that the separable convolution layer is almost 
identical to an ``extreme'' version of the inception module, which is the 
backbone of the GoogLeNet architecture \citep{googlenet}.
The Xception architecture prescribes the number of convolution kernels and output channels in each layer as well as the connections between layers.
Some key differences between our model and the standard Xception architecture 
include a batch normalization layer in front of the network, an extra 
SE layer after every residual block in the middle portion of the architecture, 
and the addition of a dropout layer at the
end 
of the network, which we 
included in place of the L2 weight regularization used in the original Xception 
model \citep{xception}.
\added{The addition of such layers is frequent practice in ML.  We included only those layers that yielded the best classification score during our hyperparameter tuning process.} 

We trained our final model for 25 epochs, where each epoch was a full pass of 
all 1,000,000 samples in the training data through the neural network. Model
loss was calculated using binary cross-entropy, which is a standard loss
function for binary classification problems that measures how well the 
predicted class probabilities match the target class. The
model performance was monitored by calculating performance metrics 
(Section~\ref{subsec:model_perf}) using the full train--dev and validation sets at the 
end of every epoch. The training was carried out on a single ML-enabled 
graphical processing unit (GeForce RTX 2060 SUPER 8 GB GPU) and took 
approximately 112 hours ($\sim4.5$ days) days to complete. 

All of our models were implemented using TensorFlow \citep{tf},
and the source code to reproduce our final model is available 
\replaced{\href{https://github.com/UCLA-SETI-Group/doom}{online}}{
online\footnote{\url{https://github.com/UCLA-SETI-Group/doom/releases/tag/v1.0.1}}
\citep{doom}}.

\section{Results}\label{sec:results}

\subsection{Model Evaluation} \label{subsec:model_eval}
Although we allowed some tolerance on the
calculation of the extrapolated detection frequencies 
in the training set (Section~\ref{subsubsec:proc_selected_signals}), we 
found that a portion of the RFI signals in the validation set were
still misclassified as valid technosignature candidates 
because the signal in the bottom image was not properly 
centered. Similarly, we found that a subset of signals were
misclassified because the S/N difference between the top and bottom images
was too large.  These discrepancies are not surprising because our
training data were generated by splitting a single scan into two parts,
while our test data contain signals from two completely different
scans. As a result, errors on the
extrapolated detection frequencies
as well as
the S/N variability are not as pronounced in the training data as they
are in the test data.

In order to address these issues, we applied several additional 
steps when evaluating the model on 
the validation, test, and production data.
First, we evaluated the model multiple times for each
image pair in the validation and test sets, applying a pixel shift in
the range -4 to 4 to the bottom image each time.  The largest of the
resulting 9 values was chosen as the score for that data point.  The
range of pixel shifts used for this step was chosen by running this
test with a larger set of pixel shift values and choosing a
symmetric range that 
yielded the largest scores for 
$\sim$95\% of the validation data.
We found that this step increased %
the validation recall from 0.859 to 0.942 for a total decrease 
of less than $0.1\%$ in the validation set precision.
Then, if the score after this step was still below the 
decision threshold value of 0.5,
we 
rescaled both images so that the new pixel values ranged from zero to the 
average of the maximum pixel values of both images prior to scaling. 
We found that this step further increased %
the validation recall from 0.942 to 0.992, while retaining a 
validation precision of $>99\%$. 

\subsection{Model Performance} \label{subsec:model_perf}

\begin{figure}[hbt]
\begin{center}
    \includegraphics[width=4in]{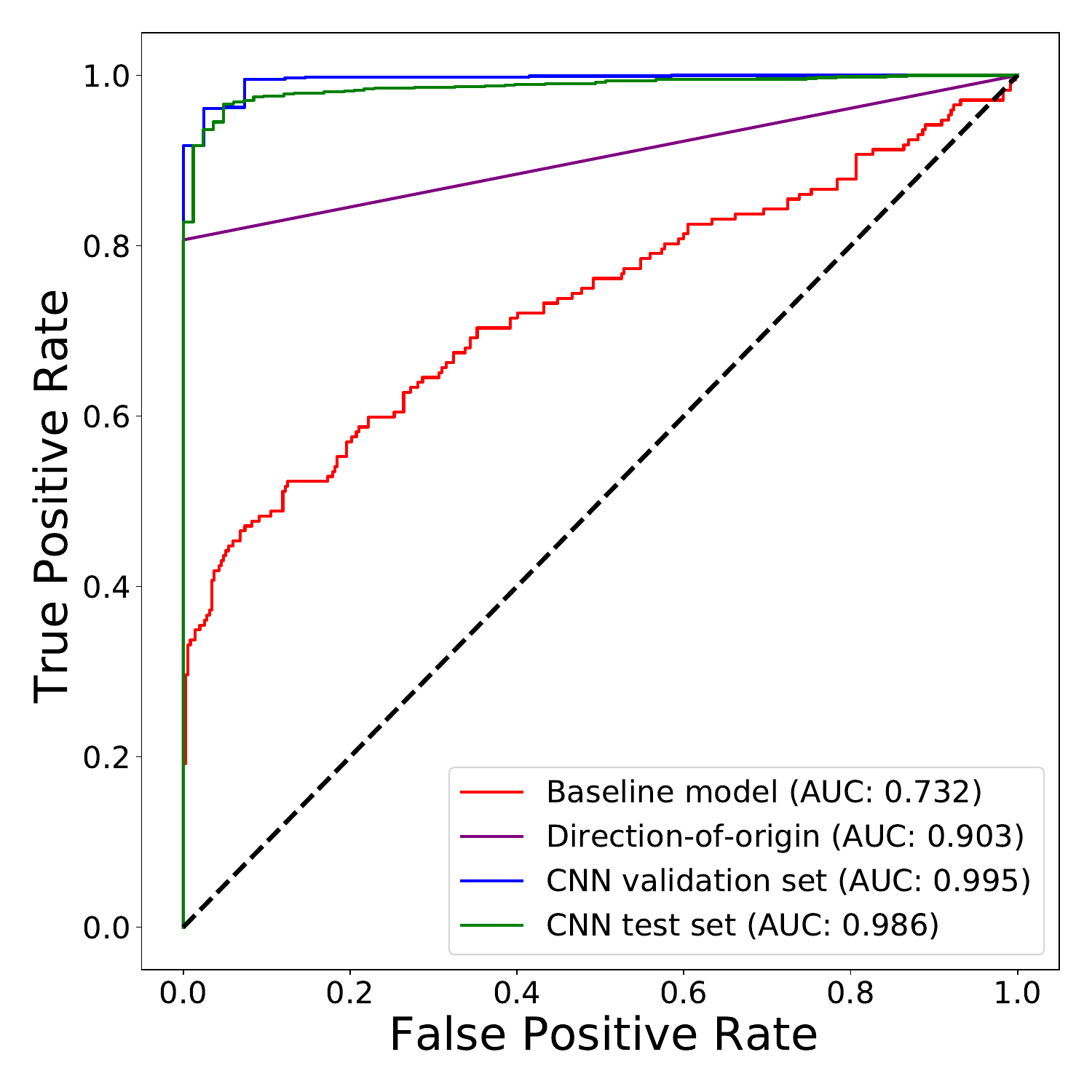}
\caption{ROC curve and AUC scores for the baseline 2D correlation model, the 
existing DoO filter, as well as the CNN evaluated on both the 
validation and the test set. The DoO curve is linear because
the filter only outputs binary scores of 0 or 1, unlike the other models, which 
output a score in the range from 0 to 1 for each sample.
The dashed line shows the ROC curve for a purely random classifier with an AUC score of 0.5. 
}
\label{fig:ROC_curves}
\end{center}
\end{figure}

We used several metrics to evaluate the performance of our model.
First, we evaluated the precision and recall scores, which are defined
in Section~\ref{subsec:im_corr}. We also calculated the ${\rm F}_1$ score, 
which is defined as 
\begin{equation}\label{eq:f1_score}
{\rm F}_1 = 2 \times \frac{P \times R}{P + R},
\end{equation}
where $P$ and $R$ are the precision and recall, respectively.
Another important metric often used for model performance evaluation is the 
\textit{area under the curve} (AUC) score. In this
context, the ``curve'' is the \textit{receiver operating characteristic} (ROC)
curve, which plots the true-positive rate against the false-positive rate
(Figure~\ref{fig:ROC_curves}).
We also calculated the area under the precision-recall curve (AUPRC) as 
well as the average precision (AP) of the model, 
which is the precision averaged over all recall values.
Together, these metrics offer a thorough
picture of the
performance of all 
models considered in this work. Table~\ref{tab:result_scores} lists the values
of these metrics for the baseline model, the existing DoO 
filter, and our trained CNN.

\begin{deluxetable}{l|cccc}[h!]
\caption{%
Scoring metrics for the baseline model, DoO filter, and CNN validation and test sets.
\label{tab:result_scores}}
\tablehead{
\colhead{Metric} & \colhead{Baseline Model} & \colhead{Direction-of-origin} & \colhead{CNN Validation} & \colhead{CNN Test}}
\startdata
Precision   &  0.9508  &  1.0000  &  0.9973  & 0.9915  \\
Recall      &  0.3372  &  0.8068  &  0.9919  & 0.9781  \\
${\rm F}_1$ &  0.4979  &  0.8931  &  0.9946  & 0.9848  \\
AUC         &  0.7324  &  0.9034  &  0.9951  & 0.9811  \\
AUPRC       &  0.6914  &  0.9975  &  0.9998  & 0.9982  \\
AP          &  0.6921  &  0.9950  &  0.9998  & 0.9991  \\
\enddata
\end{deluxetable}

We find that our CNN significantly outperforms the baseline model, with a
99.15\% precision at a recall of 97.81\%, compared to a baseline model precision 
of 95.08\% at a recall of only 33.72\%. The CNN also performs favorably with 
respect to the existing DoO filter. Although the 
DoO filter did not admit any false positives (100\% precision)
over the set of 1238 hand-labeled signals (Section~\ref{subsec:def_data_set}), 
its recall scored at only 80.68\%. This translates to a significant portion 
of signals left over for manual inspection after application of the filter.
The balance between precision and recall can be summarized with the ${\rm F}_1$,
AUC, or AUPRC scores, all of which favor the CNN over the baseline model or 
direction of origin filter. A visualization of the performance differential 
between the baseline model, the DoO filter, and the CNN is
exemplified in Figure~\ref{fig:ROC_curves}, which plots the ROC curve for each 
model.

We note that $\sim$36\% of the misclassifications on the test set are 
attributable to signals detected in scans of TRAPPIST-1 and LHS 1140. This 
percentage is appreciably larger than the total fraction of signals from these 
two sources in the full test set ($\sim$23\%). This disproportionate 
distribution may be due in part to
the much 
larger sky separation between this source 
pairing
($26.3^{\circ}$)
compared to
the typical angular separations between almost all other source pairings (1--$3^{\circ}$), including  
the other
source pairings in this observation run.
Specifically, the
time between the TRAPPIST-1 and LHS~1140 data acquisitions is approximately 
twice as large as the mean
time between data acquisitions of the other source pairs,
a unique circumstance that was driven by the desire to observe
noteworthy exoplanets that had been recently discovered at the time.
\added{Although the RFI environment is expected to be stable on these time scales}, the increased
time between data acquisitions of these two sources accentuates any
errors in the frequency extrapolation
that we used to center the
signals in the second image. These errors account for a major failure
mode discussed
in Section~\ref{subsubsec:data_failure_modes}.
The 
increased
error rate of the large-separation pairing suggests that 
such pairings
should be
avoided
when practical.
This 
information can 
help guide
the design of future observing plans.

\subsection{Application to Observational Data} \label{subsec:cnn_to_data}
We applied the trained model to a subset of the data presented by
\citet{Margot2018}, \citet{Pinchuk2019}, and \citet{Margot2021}.
Specifically, we evaluated the model on signals that passed the
drift rate, S/N, and bandwidth data selection filters described in 
Section~\ref{subsec:data_select} as well as the existing DoO filter.
Table~\ref{tab:performance_comparison} shows the total candidate signal
counts (from the first scan of each source only) before and after
application of 
the direction-or-origin filter and the CNN-based filter described in this work. 
Although the existing DoO
performs remarkably well already, we found that the CNN can further
reduce the number of signals left over to examine by a factor of
6--16 in nominal situations.  In the atypical data set with unusually
large angular separations between sources, the reduction factor decreased to $\sim$3.

\begin{deluxetable}{lrrrr}[h!b]
\caption{Comparison of filter performance across different data sets. The 
  ``Data Set'' column
  includes the
  journal and online data references for the data sets. 
The ``Total Signals'' column lists the total number of signals from the first 
scan of each source. The ``Applicable Signals'' column 
lists the total number of signals from the first scan of each source with an S/N 
between 10 and 600, a drift rate in the range $\pm2$ Hzs$^{-1}$, and a
bandwidth with FWHM $\leq 100$ Hz. The ``DoO'' column lists the number of 
candidate signals remaining after application of the existing DoO 
filter to the subset of signals from the ``Applicable Signals'' column. 
The final column lists the candidate technosignature counts after 
application of the CNN-based filter described in this work to the subset of 
signals that passed the existing DoO filter. 
\label{tab:performance_comparison}}
\tablehead{
\colhead{Data Set} & \colhead{Total Signals} & \colhead{Applicable Signals} & \colhead{DoO}  & \colhead{DoO + CNN}}
\startdata
UCLA search 2016 \citep{Margot2018, seti16dataset}  &   2,230,659 &   2,142,964  &  16,168  &  2,772  \\  %
UCLA search 2017 \citep{Pinchuk2019, seti17dataset} &   2,973,499 &   2,888,766  &  62,301  &  20,560 \\  %
UCLA search 2018--9 \citep{Margot2021, seti18dataset}  &  12,779,984 &  12,438,375  &  21,978  &  1,357  \\  %
\enddata
\end{deluxetable}

We did not evaluate the CNN on the 3\% of signals that did not pass our data 
filters (Section~\ref{subsec:data_select}) because we anticipate poor 
classification performance on these signals as they are outside of the domain of 
applicability for the network.

The evaluations took approximately 5, 18, and 6 hours on the
\citet{Margot2018}, \citet{Pinchuk2019}, \citet{Margot2021} data sets,
respectively, on a single ML-enabled GeForce RTX 2060 SUPER 8 GB
graphical processing unit,
i.e., several times slower than data acquisition.
This performance is promising with respect to near-real-time data
processing: the evaluation could keep up with data acquisition with the 
addition of one or more high-end graphical processing units.

\section{Discussion}\label{sec:discussion}

\subsection{Failure Modes} \label{subsec:failure_modes}

In this section, we examine some of the CNN failure modes that we identified by
examining the test set samples that the CNN misclassified.

\subsubsection{Model-related Failure Modes} \label{subsubsec:model_failure_modes}

One set of failure modes
falls under the category of model-related 
failures, which stem from the model's inability to learn an adequate 
representation of the data and
therefore correctly classify
a subset of signal types. 

\begin{figure}[hbt]
\begin{center}
    \includegraphics[width=4in]{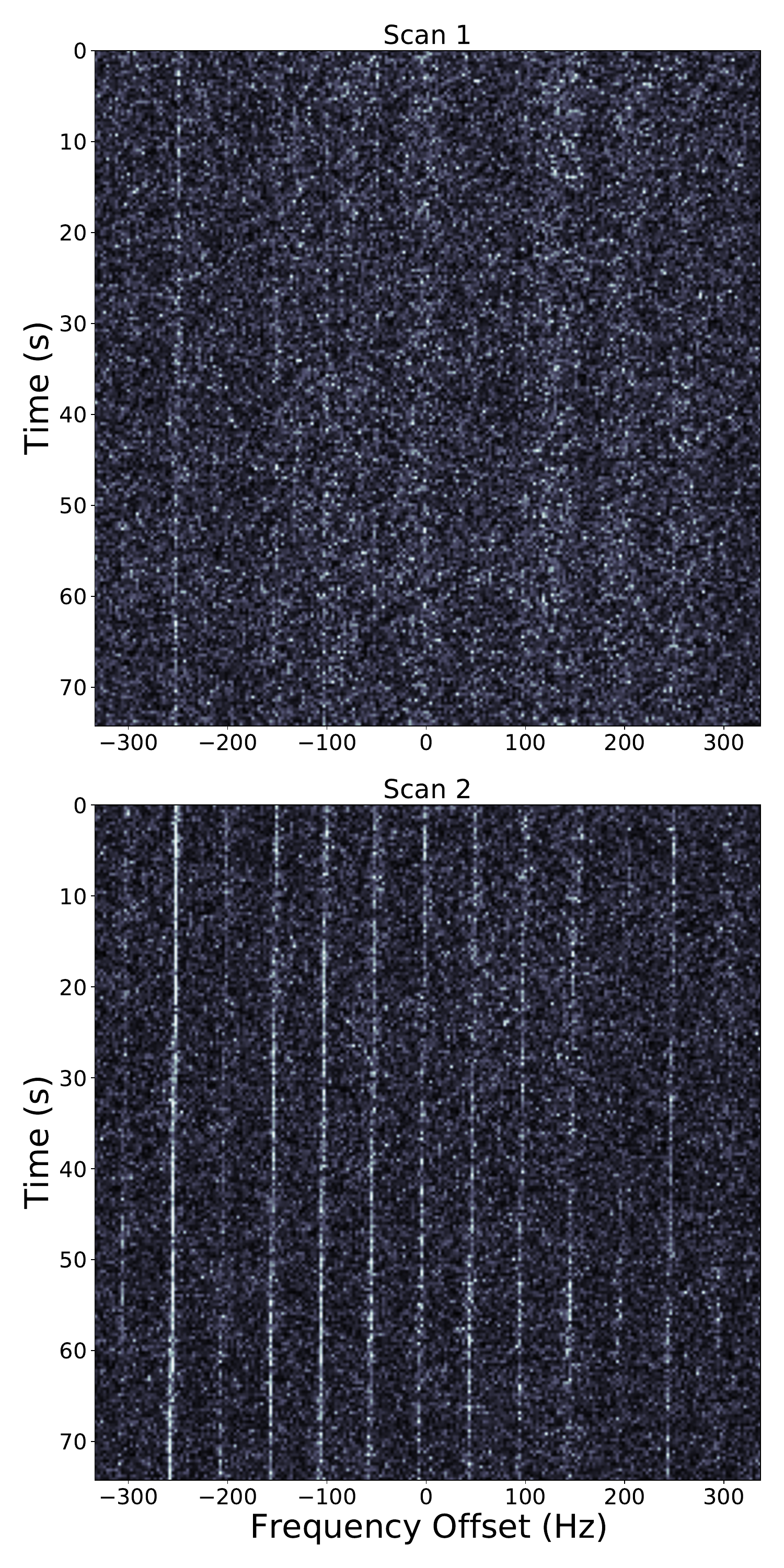}
\caption{Sample signal with lower S/N in the top image compared to the bottom image. The CNN score for this image pair is 0.2706, which corresponds to a label of ``0''.}
\label{fig:snr_fail_mode}
\end{center}
\end{figure}

One such failure mode occurred when the S/N of the signal in a 
data sample 
with S/N $<\sim30$
was lower in the top image compared to the bottom image.
In these
cases, the network would assign the pair of images a label of ``0'' when there
was clearly a signal present in both. Figure~\ref{fig:snr_fail_mode} shows an 
example of such an image pair. 
Note that we did not find any evidence for the reverse failure mode -- when 
S/N of the signal in a data sample is larger in the top image compared to the 
bottom image.
Although we did attempt to introduce S/N
variations
between the two images in each sample of the training set 
(Section~\ref{subsubsec:signal_selection}), this failure mode suggests that we needed
to allow even larger variations, specifically including cases where the S/N
of the top image is substantially
lower than the S/N of the signal in the 
bottom image.

\subsubsection{Failure Modes Related to Simplifying Assumptions} \label{subsubsec:data_failure_modes}

\begin{figure}[hbt]
\begin{center}
    \includegraphics[width=7in]{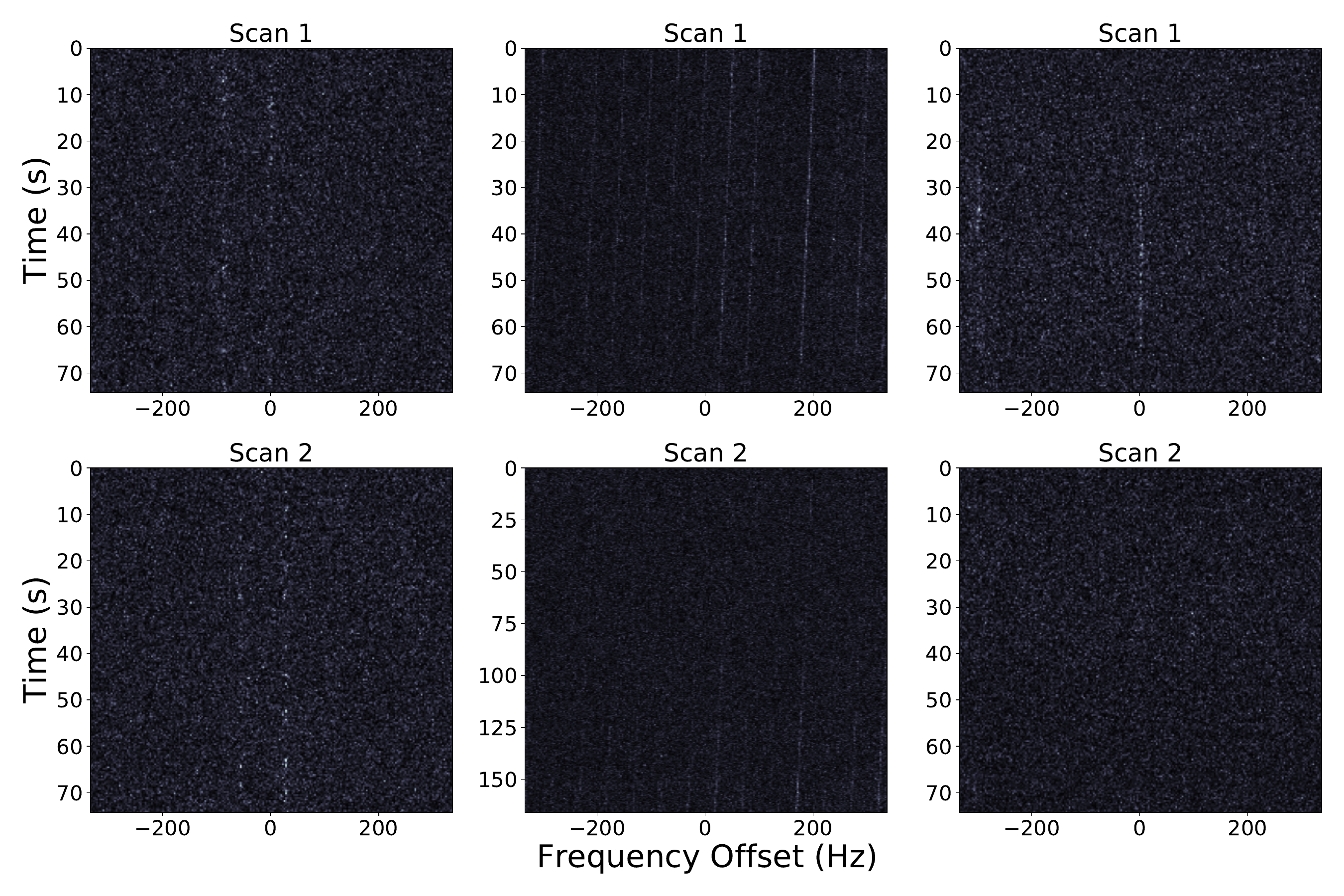}
\caption{(Left) Example image pair where the signal in the second image is not centered. This occurs when the properties of the signal in the top image are inaccurately determined. (Middle) Example of a signal that does not appear until the second half of the scan in the second image. The standard application of the network mislabels this signal because the CNN looks at the top half of each scan only. (Right) Example signal that was incorrectly hand-labeled as ``1'', seemingly indicating that it contains a signal in the second image.}
\label{fig:fail_modes}
\end{center}
\end{figure}

A different set of failure modes 
stemmed from some of the simplifying assumptions that we made about
the data.  For example, one of the failure modes
is 
related to the frequency extrapolation that we performed in order to
center the signal in the second image.
We assumed that the frequency drift rate would be linear, and we
assumed that our estimate of the frequency drift rate would be accurate enough
to ensure centering of the signal in the second image within a
tolerance of $\sim$15 Hz.  Although we
included this tolerance 
directly into the model, during both training
(Section~\ref{subsubsec:proc_selected_signals}) and evaluation
(Section~\ref{subsec:model_eval}), we still found cases where the
signal was clearly present in the second image but not properly
centered.  Figure~\ref{fig:fail_modes} (left) shows an example of such
a signal from our test set. In this case, the model gave the sample a
score of 0.0193. This score
yields a label of ``0'' (i.e., no
signal in the second scan)
because it is below the
decision
threshold of 0.5.
However, if we shift the bottom image 5 pixels 
($\sim$15 Hz) to the left, the score jumps to 0.7545, which
yields the 
correct label of ``1.'' Shifts of 6--10 pixels to the left all yield scores 
$> 0.99$. 

Unfortunately, this problem cannot be fixed by simply increasing the range of 
shifts allowed for the bottom image. In fact, it is likely that the same problem 
persists
for any choice of the tolerance on frequency to accept/reject a match.
More importantly, increasing the 
range of allowed shifts for the bottom image would also increase the risk of removing 
a technosignature candidate by pairing it with RFI detected in its vicinity.
Instead, this problem can be
better addressed by obtaining more accurate 
representations of the detected signal properties, which can then be used
to more accurately localize and thus center the signal in a subsequent scan.

Another failure mode is related to the simplifying assumption that the signal power in the first half of a scan is comparable to the signal power in the second half of a scan.
The input to the CNN is limited to approximately half of 
each scan,
with the scan parameters of \citet{Pinchuk2019},
by virtue 
of the training set parameters. Specifically, the input images are 
limited to a size of 225$\times$225 pixels,
which corresponds to $\sim$75 seconds 
of observation, whereas each scan typically lasts for a total of $\sim$150 
seconds.
This limitation significantly hinders the network's ability to 
identify RFI, because the CNN only examines the first temporal half of each scan, 
but there
are instances where a signal is only present in the latter half 
of one or both scans. An example of this case is shown in Figure~\ref{fig:fail_modes} (middle). 

We performed
three preliminary attempts at mitigating this issue. First, we tested 
the possibility of
downsampling the scan in the time dimension
by a factor of two,
effectively allowing us to fit partial information from 450 rows ($\sim$150 
seconds) into 225
pixels along the time dimension. 
Second, we tested an image rescaling approach consisting of
a linear interpolation \citep{scipy} of the entire scan duration 
that was sampled at
225 equally spaced time intervals. Neither approach reduced the number of 
signals left to examine after application of the filter, indicating that the 
issue persisted. %
In a third attempt, we applied the filter a total of four times
to
each set of
scans in the test data set.  Each filter evaluation paired a different
set of temporal scan halves
(scan$_{1,{\rm top}}$ with scan$_{2,{\rm top}}$, scan$_{1,{\rm top}}$ with scan$_{2,{\rm bottom}}$, etc.).  
We combined the
results of these evaluations by taking the maximum score across the
four trials.  We found that this method did increase the recall score
for the test set from 97.81\% up to 98.91\%. However, the precision
score was heavily penalized, decreasing from 99.15\% down to
97.92\%. 
This trade-off %
increases the likelihood of finding additional pairings and therefore false positives.
Taking the median score across the four trials yielded similar results.
This four-execution mitigation attempt
also 
increased the computational cost
of the CNN filter by a factor of four.
For these reasons, we did not apply this method when evaluating the CNN on 
observational data. 
Further investigation beyond the scope of this work is required to minimize
the impact of this failure mode.

\subsubsection{Other Failure Modes} 

The final failure mode that we observed is
related to instances of human error in 
labeling the validation and test sets. Because the labels were supplied by a 
single
classifier (PP),
the margin for error on the validation and test labels is nonzero.
Figure~\ref{fig:fail_modes} (right) shows an example of a test signal 
that the CNN ``misclassified''. Upon further investigation, it is clear that the 
label provided with this data sample is incorrect. Although the network 
technically classified this signal correctly, it counted as a misclassification 
when computing model performance (Section~\ref{subsec:model_perf}). This problem 
could be substantially mitigated if multiple
people examined and labeled the validation and test data.

\subsection{Future Improvements} \label{subsec:future_improvements}
Though we have attempted to thoroughly search the parameter space for
the best model to perform our classification task, there are still a number
of options to consider for future improvements. For example, for all 
CNN models considered in this work, the input was comprised of
225$\times$225$\times$2 images, where the last dimension distinguished the 
top half of the first scan from the top half of the second scan.
An alternative approach would pass the data as a 
single 450$\times$225 image, where the top
halves of each scan
are concatenated in 
the time dimension. It is worth investigating whether or not this variant on the 
input data improves network classification performance. Along the same lines, it 
may be beneficial to train a denoising autoencoder \citep[e.g.,][and references 
therein]{DAE} and apply it to the images prior to sending them through the 
CNN. If the denoising autoencoder functions properly
(i.e., reduces
the noise around the signals in the image),
it is likely that the
CNN would receive a
boost in classification performance.

During our model selection step (Section~\ref{subsec:model_selection}), 
we found that standard network architectures always outperformed their
Siamese variants. However, those tests were performed without any 
hyperparameter tuning, so it may be worthwhile to investigate whether
a tuned Siamese model still underperforms when compared to the base 
architecture model.
On top of that, new state-of-the-art CNN architectures are still being rapidly 
developed and may offer significant improvements over the Xception 
architecture used as the final model in this work.
For example, the novel EfficientNet architecture \citep{tan2019efficientnet},
which was published after our model selection efforts,
is almost an order of magnitude smaller and faster than other CNN 
architectures, yet has been shown to exhibit state-of-the-art performance on the 
ImageNet data.

Finally, there are some improvements that can be made to the overall training 
and evaluation process to mitigate the various failure modes discovered after 
evaluating the CNN on the test data. These improvements are included with the 
corresponding description of each failure mode in Section~\ref{subsec:failure_modes}.

\section{Conclusions}\label{sec:conclusions}

In this work, we designed a DoO filter using modern computer
vision techniques to assist in the mitigation of RFI
in the search for radio technosignatures.
We began by
randomly selecting 1,100,000 signals from a
carefully selected set of over 8 million detections in order to 
obtain the cleanest training and train-dev data sets possible. Both of these data 
sets consist of pairs of images that were obtained by splitting a single scan 
containing a signal into two parts. This approach allowed us to label a large 
number of signals in a short period of time. 

Using these data sets, we trained and evaluated a CNN designed to determine 
whether or not the signal in the first image is also present in the second image. 
This network can therefore be applied
to determine if
a detected signal is persistent in one and only one direction on the sky. This 
approach is similar to the one employed by
traditional DoO filters, 
except that the CNN analyzes the dynamic spectra directly instead of relying on 
inferred signal properties,
such as frequency and frequency drift rate.

We found that the CNN trained in this work outperformed both the baseline 2D 
correlation model and the existing DoO filters, with a precision 
value of 99.15\%  at a recall of 97.81\%. We find that the CNN can reduce the 
number of signals left to analyze after applying the existing DoO 
filter by a factor of
6--16 in nominal situations.  In the atypical data set with unusually
large angular separations between sources, the reduction factor decreased to $\sim$3.

We identified several failure modes of the trained network,
labeling failures, and failures related to simplifying assumptions.
Each failure mode can be 
addressed 
with future CNN versions 
to increase
the classification 
performance. 
Integrating this ML-based DoO filter into existing radio technosignature
search pipelines has the potential of providing accurate RFI identification in near-real time.

\acknowledgments

PP thanks David Saltzberg, Troy A.\ Carter, 
and Michael P.\ Fitzgerald for useful discussions. JLM thanks Tuan Do for useful discussions.
PP and JLM were supported in part by NASA grant 80NSSC21K0575.  PP was funded in part by the Nathan P.\ Myhrvold
Graduate Fellowship.
\added{Funding for the UCLA SETI Group was provided by The Queens Road
Foundation, Janet Marott, Michael W. Thacher and Rhonda L. Rundle,
Larry Lesyna, and other donors (\url{https://seti.ucla.edu}).}
\software{TensorFlow \citep{tf},  
NumPy \citep{numpy},
SciPy \citep{scipy},
scikit-learn \citep{sklearn},
pandas \citep{pandas},
Matplotlib \citep{mpl}
}

\appendix

\section{Sample training signals} \label{app:example_sigs}

\begin{figure}[htb]
  \begin{center}
    \includegraphics[width=7in]{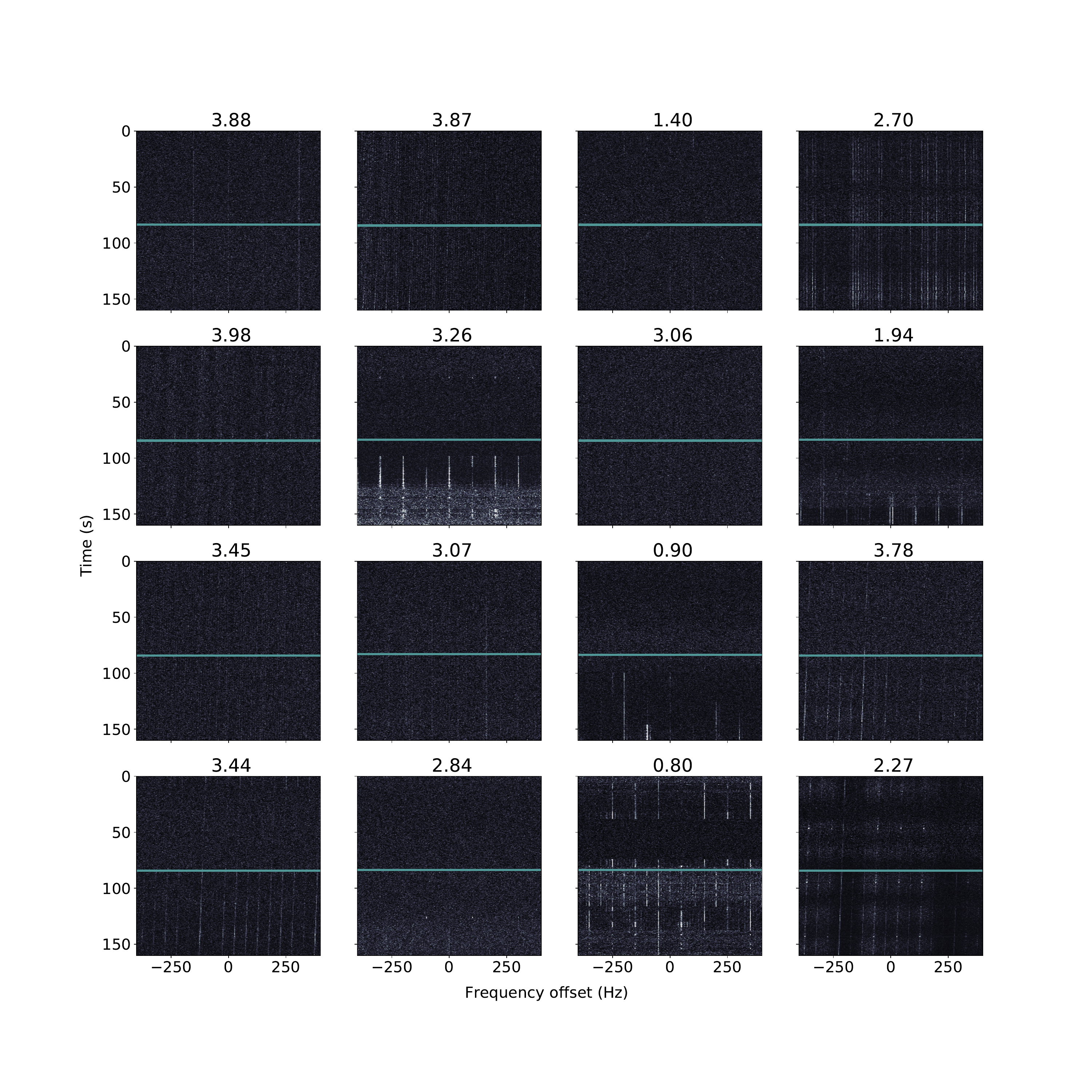}
    \caption{Example dynamic spectra
      of signals with a top-half S/N $<6$ (exact values shown above each plot). The horizontal blue line delimits the top and bottom halves. Note that these signals (located in the center of the image 
starting at 0 Hz offset at time $t=0$) 
are faint in the top half of the image and difficult to detect visually.
    }
\label{fig:top_half_snr_low_examples}
\end{center}
  \end{figure}

\begin{figure}[htb]
  \begin{center}
    \includegraphics[width=7in]{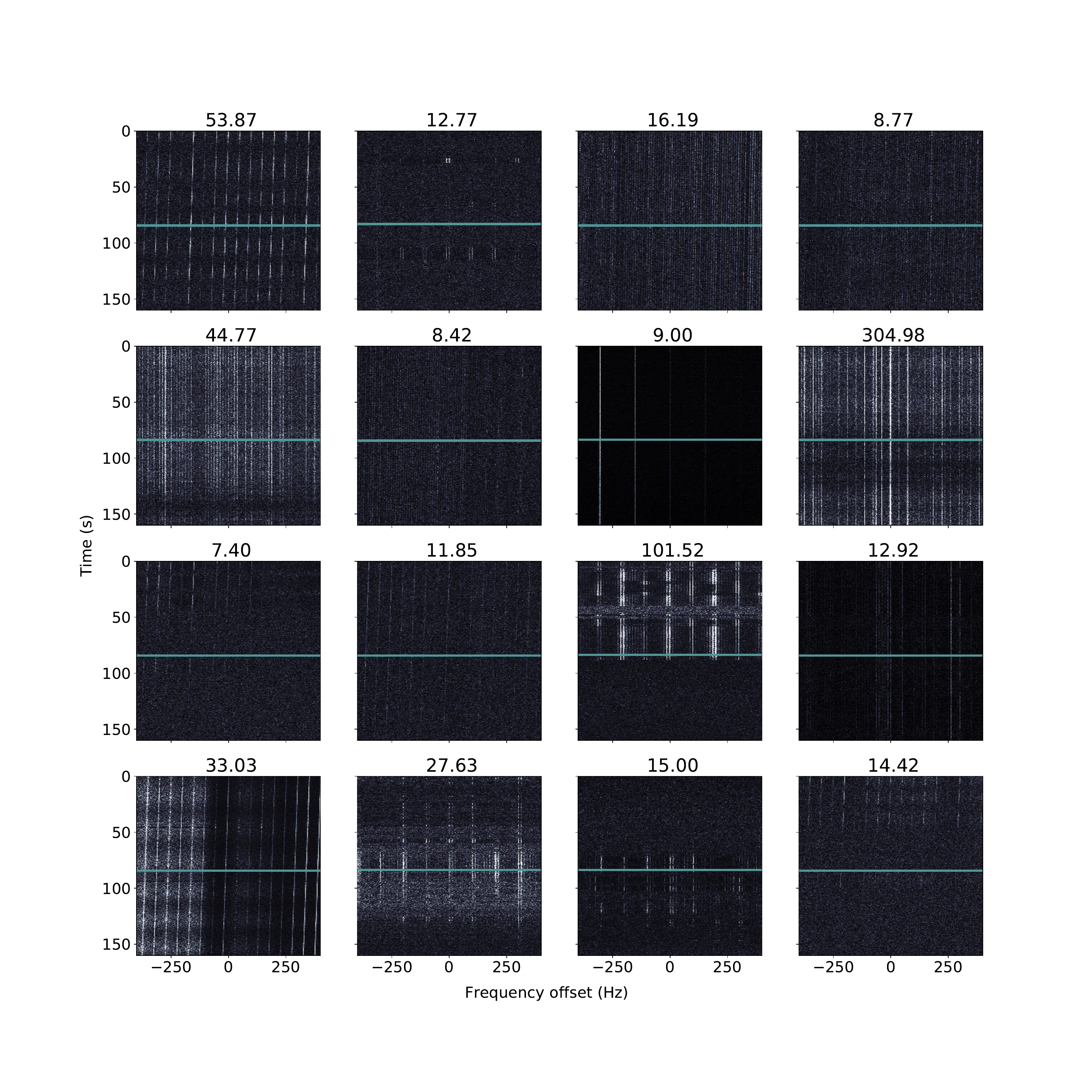}
    \caption{Example
      dynamic spectra of signals with a top-half S/N $\geq 6$ (exact values shown above each plot). The horizontal blue line delimits the top and bottom halves. Note that all of these signals (starting at 0 Hz offset at time $t=0$) are visually detectable in the top
      half
of each sample.}
\label{fig:top_half_snr_high_examples}
\end{center}
  \end{figure}

We validated our choice for the threshold
of a top-half S/N $<6$
(Section~\ref{subsubsec:signal_selection})
by examining a sample of signals below this cutoff value.
Figure~\ref{fig:top_half_snr_low_examples} depicts these signals,
most of which are faint and difficult to
detect visually.
Similarly, Figure~\ref{fig:top_half_snr_high_examples} depicts a sample of 
signals above this threshold, which are clearly visible in the
dynamic spectra.
  
\begin{figure}[htb]
  \begin{center}
    \includegraphics[width=7in]{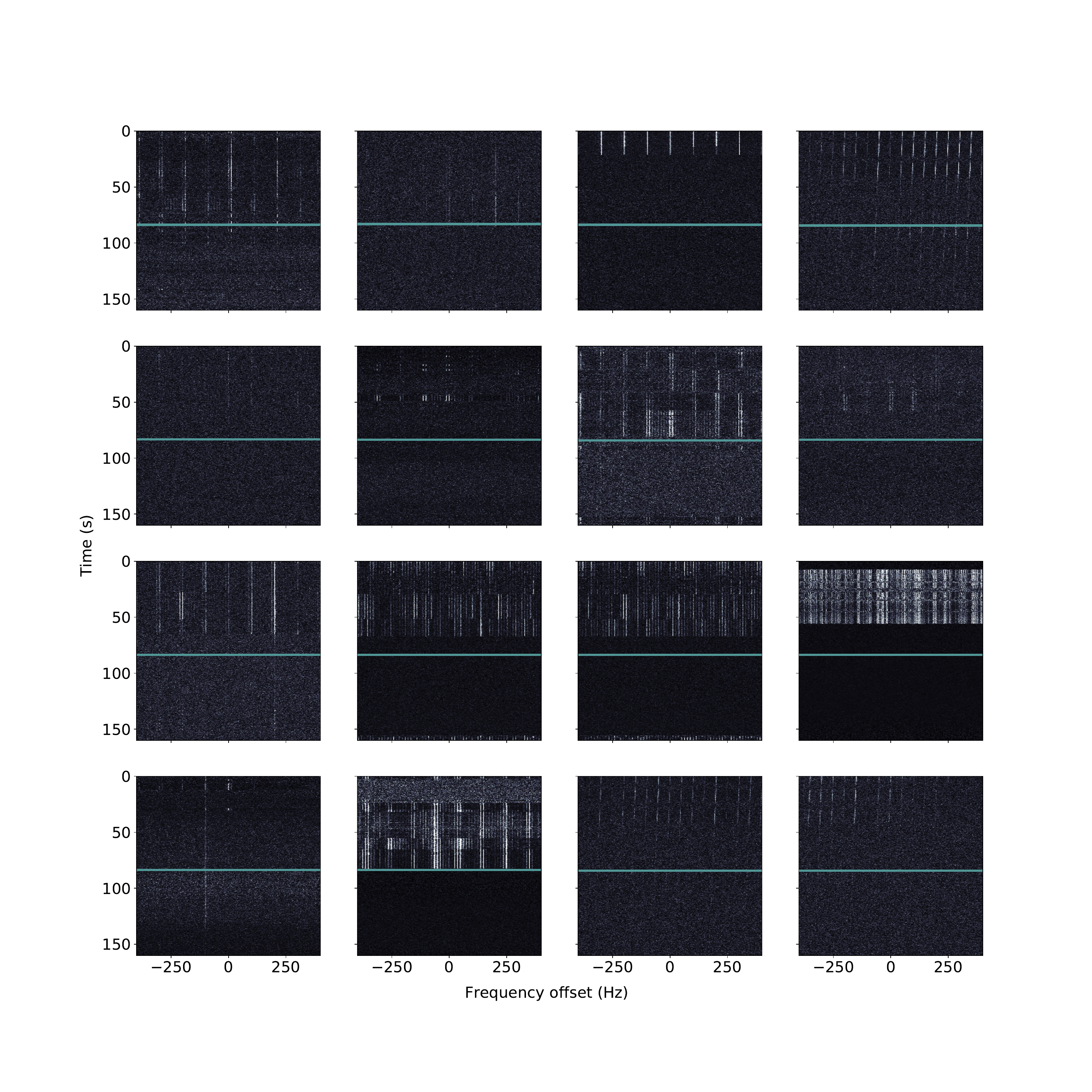}
\caption{Example time-frequency diagrams of signals with a $P_{\rm bottom} / P_{\rm top}$ ratio of 0.2 or lower. The horizontal blue line delimits the top and bottom halves. No signals with a prominence value greater than $3\sigma$ are present in the bottom halves. All of these signals represent valid negative samples.}
\label{fig:pure_neg_sig_examples}
\end{center}
  \end{figure}

Figure~\ref{fig:pure_neg_sig_examples} shows a sample of signals from the 
negative category with a $P_{\rm bottom} / P_{\rm top}$ ratio of 0.2 or lower and 
prominence value below
3 standard deviations of the noise (Section~\ref{subsubsec:signal_selection}).

\section{Siamese models} \label{app:siamese_models}

The concept of a ``Siamese'' neural network was first introduced in 
1993 by \citet{siamese1} for the purpose of signature verification.
Siamese networks are defined as two identical subnetworks that are joined at 
the output, typically by subtracting the neuron values of the final layer of 
one model from the neuron values of the final layer of the other model. 
The input to these networks always consists of two data points, each of which 
are passed to one of the two subnetworks. The output of the Siamese networks 
is typically given as a similarity score between the two data points.

\begin{figure}[htb]
  \begin{center}
    \includegraphics[width=7in]{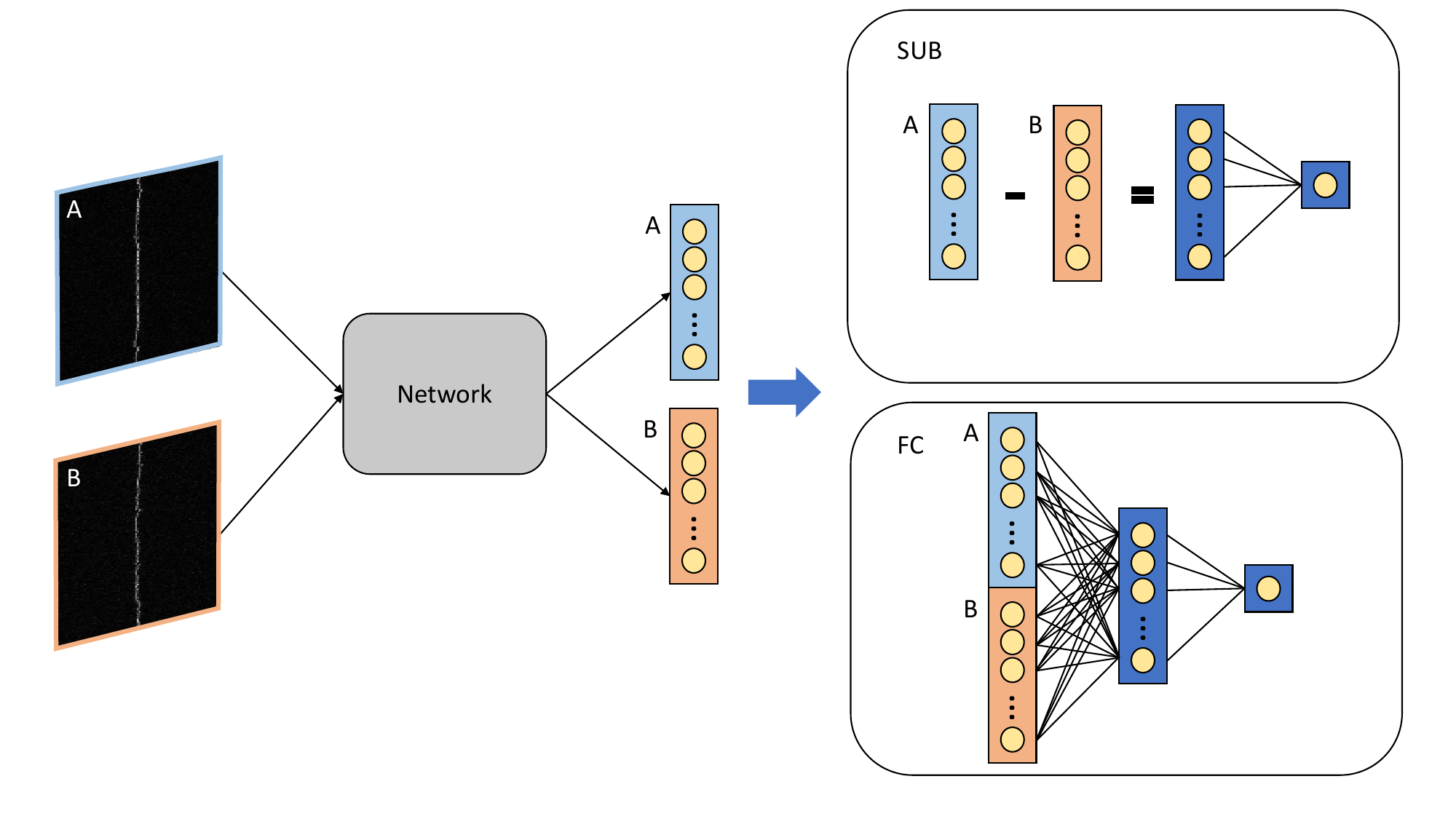}
\caption{Example of a Siamese network tested in this work. 
The labels ``A'' and ``B'' represent input from two different scans. The 
``Network'' in the middle was replaced with one of the architectures that was 
tested in this work (see Section~\ref{subsec:model_selection}). The output
layers were joined in two different ways: (Top Right) In standard Siamese 
networks, the output layers are subtracted. (Bottom Right) In our generalized
version, the output layers are concatenated and connected to another layer 
with $N$ neurons. Equation~\ref{eq:siamese_weights} gives the set of weights 
for this configuration that reproduce the standard layer subtraction 
procedure.}
\label{fig:siamese_model_architecture}
\end{center}
  \end{figure}

During the model selection portion of this work (see Section~\ref{subsec:model_selection}), 
we set the two identical subnetworks of each 
Siamese network to be one of the architectures under 
consideration (Figure~\ref{fig:siamese_model_architecture}). 
Each subnetwork received one scan as input.
For each architecture, we tested two methods of joining the outputs of the 
final layers of the Siamese subnetworks. Specifically, we considered the 
standard method of subtracting the values of one output from the other, as 
well as a generalized version of this procedure. For the latter, we 
concatenated the output weights of both subnetworks and added another fully 
connected layer with $N$ nodes immediately after the concatenated layer, where 
$N$ is the number of neurons in the output layer of the subnetwork.
This method is a generalized version of the subtraction procedure because
it can be recovered by setting the weights $w_{ij}$ between the two layers to 
be
\begin{equation}\label{eq:siamese_weights}
w_{ij} = 
\begin{cases}
    1,   & \text{if } i = j\\
    -1,  & \text{if } i = j+N\\
    0,   & \text{otherwise}
\end{cases}
\end{equation}
where $i$ and $j$ represent the indices of the neurons of the concatenated 
and fully connected layers, respectively.

Although Siamese networks seem like a promising solution to the problem of 
pairing signals from two different scans, we found that the standard network 
architectures always outperformed their Siamese variants 
(see Section~\ref{subsec:model_selection}). 

\pagebreak
\bibliography{setiML}

\listofchanges

\end{document}